\documentclass[12pt,preprint]{aastex}
 \usepackage{url}

\shorttitle{The energy partition of GRB/SN events}
\shortauthors{L\"{u} et al.}
\slugcomment{}
\begin{document}

\title{Gamma-Ray Burst/Supernova Associations: Energy partition and the case of a magnetar central engine}

\author{Hou-Jun L\"{u}\altaffilmark{1}, Lin Lan\altaffilmark{1},
Bing Zhang\altaffilmark{2,1}, En-Wei Liang\altaffilmark{1}, David Alexander Kann\altaffilmark{3},
Shen-Shi Du\altaffilmark{1}, and Jun Shen\altaffilmark{1}} \altaffiltext{1}{Guangxi Key Laboratory for
Relativistic Astrophysics, Department of Physics, Guangxi University, Nanning 530004, China;
lhj@gxu.edu.cn}\altaffiltext{2}{Department of Physics and Astronomy, University of Nevada Las Vegas,
Las Vegas, NV 89154, USA; zhang@physics.unlv.edu}\altaffiltext{3}{Instituto de Astrof\'{\i}sica de
Andaluc\'{\i}a (IAA-CSIC), Glorieta de la Astronom\'{\i}a s/n, E-18008 Granada, Spain}

\begin{abstract}
The favored progenitor model for Gamma-ray Bursts (GRBs) with Supernova (SN) association is the core
collapse of massive stars. One possible outcome of such a collapse is a rapidly spinning, strongly
magnetized neutron star (``magnetar''). We systematically analyze the multi-wavelength data of GRB/SN
associations detected by several instruments before 2017 June. Twenty GRB/SN systems have been
confirmed via direct spectroscopic evidence or a clear light curve bump, as well as some spectroscopic
evidence resembling a GRB-SN. We derive/collect the basic physical parameters of the GRBs and SNe,
and look for correlations among these parameters. We find that the peak brightness, $\rm ^{56}Ni$ mass,
and explosion energy of SNe associated with GRBs are statistically higher than other Type Ib/c SNe. A
statistically significant relation between the peak energy of GRBs and the peak brightness of their
associated SNe is confirmed. No significant correlations are found between the GRB energies (either
isotropic or beaming-corrected) and the supernova energy. We investigate the energy partition within
these systems and find that the beaming-corrected GRB energy of most systems is smaller than the SN
energy, with less than 30\% of the total energy distributed in the relativistic jet. The total energy
of the systems is typically smaller than the maximum available energy of a millisecond magnetar
($2\times 10^{52}$ erg), especially if aspherical SN explosions are considered. The data are consistent
with-though not proof of-the hypothesis that most, but not all, GRB/SN systems are powered by
millisecond magnetars.

\end{abstract}


\section{Introduction}
Gamma-ray bursts (GRBs) and supernovae (SNe) are known as the brightest and most powerful explosions in
the universe, with a typical isotropic emission energy of $\sim10^{52}$ and $\sim10^{51} ~\rm erg$,
respectively (Woosley \& Bloom 2006; Hjorth \& Bloom 2012; Kumar \& Zhang 2015; Cano et al. 2017b).
Despite the similarity in the released energy between these two types of phenomena, a direct connection
between them was not established until the discovery of the first association between an under-luminous
GRB 980425 and a Type Ic SN 1998bw at redshift $z=0.0085$ (Galama et al. 1998; Kippen et al. 1998; Pian
et al. 1998; Sadler et al. 1998). A handful of long GRBs associated with spectroscopically identified
SNe were henceforth detected, e.g., GRB 030329A/SN 2003dh (Hjorth et al. 2003; Stanek et al. 2003;
Kovacevic et al. 2014). More generally, long GRBs typically occur in active star-forming regions in
irregular star forming galaxies (Fruchter et al. 2006). All these suggest a direct connection between
long GRBs and a special type of Type Ic SNe, both of which are related to the collapse of a special
type of massive stars (likely the so-called Wolf-Rayet stars) known as the ``collapsars'' (e.g.,
Woosley 1993; Paczy\'{n}ski 1998; Woosley \& Bloom 2006; Cano et al. 2017b).

In general, asymmetric stellar explosions invoke a central engine to power the supernova and possibly a
GRB (e.g., Bisnovatyi-Kogan 1970). Two types of post-collapse central engine models have been discussed
in the literature for these explosions (e.g., Kumar \& Zhang 2015, Zhang 2018 for a review): one
invoking a stellar-mass black hole fed by an accretion disk (e.g., Popham et al. 1999; Narayan et al.
2001; Lei et al. 2009; van Putten et al. 2011; Liu et al. 2017), and the other invoking a rapidly
spinning, strongly magnetized neutron star called a magnetar (Usov 1992; Thompson 1994; Dai \& Lu
1998a,b; Wheeler et al. 2000; Zhang \& M\'esz\'aros 2001; Metzger et al. 2011; Bucciantini et al. 2012;
L\"{u} \& Zhang 2014).

From the observational point of view, evidence for a magnetar central engine has been collected in both
GRBs and SNe. In the {\em Swift} era, a good fraction of both long and short GRBs exhibit an X-ray
plateau followed by a very sharp drop with a temporal decay slope steeper than three, which is known as
an internal plateau. This feature is difficult to interpret by the external shock model or by the
models invoking a black hole central engine, but it is consistent with the internal dissipation  of a
long-lasting jet launched by a spinning-down magnetar, which collapses into a black hole at the end of
the plateau (e.g., Troja et al. 2007; Lyons et al. 2010; Rowlinson et al. 2010, 2013; L\"u \& Zhang
2014; L\"{u} et al. 2015; De Pasquale et al. 2016a). On the other hand, the so-called super-luminous
SNe (SLSNe), which have a luminosity tens of times higher than normal core-collapse supernovae, are now
being routinely detected (Quimby et al. 2007; Gal-Yam 2012; Nicholl et al. 2015). At least some of them
require additional energy injection to power the SN emission (Quimby et al. 2011; Nicholl et al. 2014;
Wang et al. 2015). The magnetar model is a viable possibility to explain these events by providing the
rotational energy via magnetic dipole radiation\footnote{Some suggested SLSNe, e.g., the most luminous
one (ASASSN-15lh or SN 2015L; Dong et al. 2016) claimed so far, have been also explained in terms of
models other than the magnetar model, e.g., tidal disruption events on to a Kerr black hole (Leloudas
et al. 2016; Kr{\"u}hler et al. 2018) or the spin-down of a stellar Kerr black hole (van Putten \&
Della Valle 2017).} (e.g., Kasen \& Bildsten 2010; Woosley 2010; Dessart et al. 2012; Nicholl et al.
2014; Vreeswijk et al. 2014; Metzger et al. 2015; Wang et al. 2015, 2016; Yu et al. 2017). Mazzali et
al. (2014) noticed that the kinetic energy of SNe associated with GRBs tends to cluster near $10^{52}$
erg, which is below the maximum magnetar spin energy. They then suggested that GRB-SNe may be powered
by underlying magnetars.

In any case, magnetars are likely operating in at least some super-luminous SNe and GRBs. One therefore
has the following questions. Is it common to have magnetars power GRBs in general, and in particular
SN-associated GRBs? What is the energy partition in these events between the relativistic jet (prompt
and afterglow emission of the GRB) and the more isotropic emission (SN)? Are there correlations between
parameters related to GRBs and SNe?

This paper aims to address these interesting questions through a systematic analysis of a sample of
SN-GRB associations. The criteria for sample selection and the performed data analysis are presented in
Section 2. Section 3 shows some statistical comparisons of the physical properties of GRBs and SNe and
their correlations. The case of a magnetar central engine and the energy partition between GRB and SN
in our sample are studied in Section 4. The conclusions are drawn in Section 5 with some discussions.
Throughout the paper, a concordance cosmology with parameters $H_0 = 71$ km s$^{-1}$ Mpc $^{-1}$,
$\Omega_M=0.30$, and $\Omega_{\Lambda}=0.70$ is adopted to calculate the energetics of GRBs and SNe.

\section{Sample selection and data analysis}
\begin{figure}
\includegraphics[angle=0,scale=0.4]{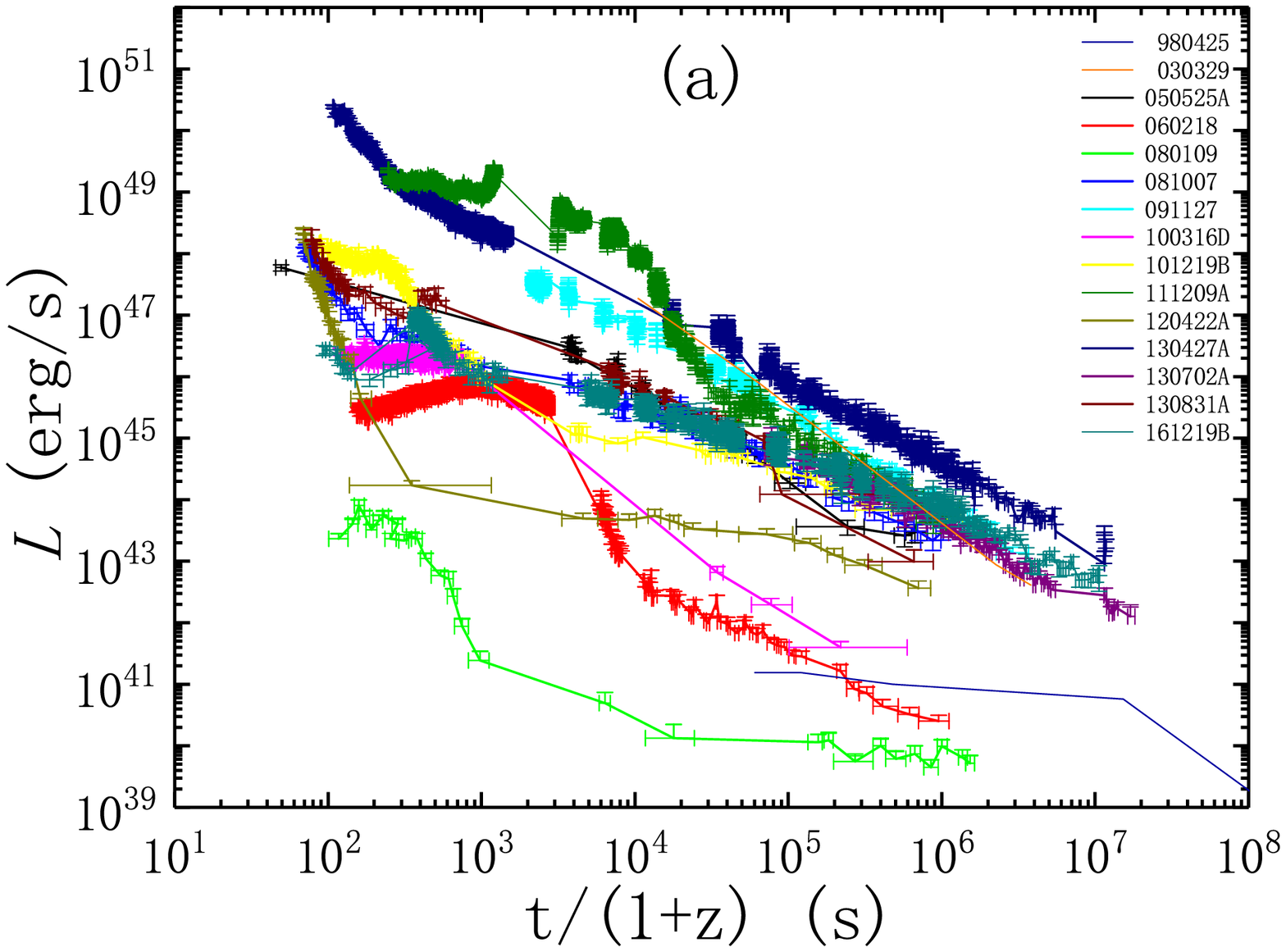}
\includegraphics[angle=0,scale=0.4]{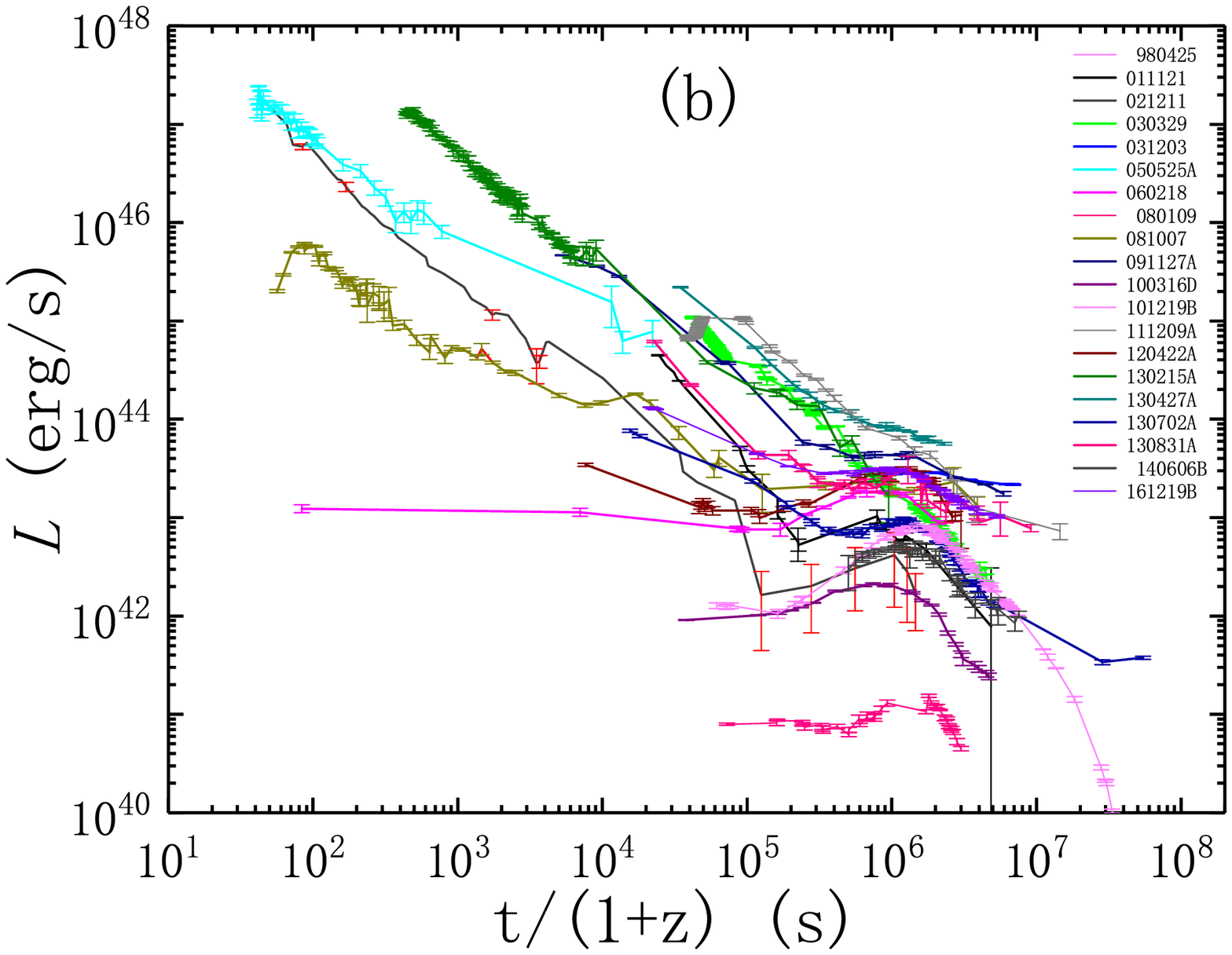}
\center\caption{The X-ray (a) and optical (b) luminosity light curves of the GRB/SN systems in our sample in
the rest frame.}
\label{fig:XRTOPT}
\end{figure}

We extensively searched for the claimed GRB/SN associations before 2017 June from the literature. The
criteria of sample selection is that either the associated SN must be confirmed via spectroscopic
evidence (SN spectral features in the optical band), or a clear light curve bump is observed at late
times in the GRB afterglow emission, and in the meantime a SN is observed independent at the same
location with spectroscopic evidence resembling a GRB/SN. To remove ambiguity, we do not include those
cases with a bump in the optical afterglow without spectroscopic support. Our entire sample includes 20
GRB/SN events. Figure \ref{fig:XRTOPT} shows the X-ray and optical light curves of these GRBs in the
rest frame\footnote{The X-ray afterglow data of GRBs 011121, 021211, 031203, 130215A, and 140606B are
missing due to observational constraints, so they are not presented in the figure.}. The redshift
(determined from the spectral lines of the host galaxies) and the GRB emission properties are collected
from the literature and presented in Table 1.

The properties of the associated SNe are presented in Table 2. The type of the GRB-associated SN is
mostly Type Ic except GRB 111209A/SN 2011kl, which was identified as a super-luminous SN (Greiner et
al. 2015; Kann et al. 2016; see Table 2). A special X-Ray Outburst (XRO) 080109 (a type of cosmological
X-ray transient due to SN shock breakout with a luminosity much lower than GRBs) is also included in
our sample, which is associated with a Type Ib SN 2008D (Soderberg et al. 2008).

\section{Statistical Properties of GRB/SN events and their Possible Correlations}
Our purpose is to compare the observed properties of our GRB/SN associations sample with other typical
long GRBs and Type Ib/c SNe, and find out the differences and similarities between them.

\subsection{Physical parameters of GRBs}
The isotropic prompt $\gamma$-ray emission energy ($E_{\rm \gamma, iso}$) of GRBs is usually derived
from the observed fluence ($S_\gamma$) in the detector's energy band, and extrapolated to the
rest-frame 1-$10^4$ keV using spectral parameters. It is given by
\begin{eqnarray}
E_{\rm \gamma,iso}&=&4\pi k D^{2}_{L} S_{\gamma} (1+z)^{-1}\nonumber \\
&=&1.3\times 10^{51}~ {\rm erg}~ k D^{2}_{L,28} (1+z)^{-1} S_{\gamma,-6},
\end{eqnarray}
where $z$ is the redshift, $D_{L} = 10^{28}~{\rm cm}~D_{L,28}$ is the luminosity distance, and $k$ is
the $k$-correction factor from the observed band to 1-$10^{4}$ keV in the rest frame (e.g., Bloom et
al. 2001; L\"{u} et al. 2014). The convention $Q = 10^x Q_x$ is adopted in cgs units for all parameters
throughout this paper. As the spectra of most GRBs in our sample can be modeled with the so-called
``Band function'' (Band et al. 1993) or the cutoff power-law model, the peak energy of spectrum
($E_{\rm p}$) can be measured from the data. Here, we do not analyze the spectra of GRBs systematically
by ourselves, but collect the $E_{\rm p}$ values from the published papers. The $E_{\rm \gamma, iso}$
of GRBs are reported in Table 3.

Another important parameter is the isotropic kinetic energy $E_{\rm K,iso}$, which is measured from the
afterglow flux if the normal decay segment of the X-ray or optical afterglow can be observed. This is
because this value becomes constant during the normal decay phase (after energy injection during the
prior shallow decay phase, Nousek et al. 2006; Zhang et al. 2006). Following the method discussed in
Zhang et al. (2007), we calculate $E_{\rm K,iso}$ based on the normal decay phase using the X-ray or
optical data. We use the ``closure relation''\footnote{This is the relation between temporal $\alpha$
and spectral $\beta$ index (Zhang \& M\'esz\'aros 2004; Zhang et al. 2006; Gao et al. 2013).} to judge
the spectral regime and the profile of the circumburst medium, i.e., (1) $\nu_m < \nu < \nu_c$ for the
interstellar medium (ISM) model; (2) $\nu_m < \nu < \nu_c$ for the Wind model; and (3) $\nu>{\rm
max}(\nu_m,\nu_c)$ for both the ISM and Wind model (in this case, the $E_{\rm K,iso}$ expression does
not depend on the medium density, Zhang et al. 2007; L\"{u} \& Zhang 2014). These derivations depend on
the unknown shock equipartition parameters for electrons ($\epsilon_e$) and for magnetic fields
($\epsilon_B$). In our calculations, we assume $\epsilon_{\rm e}=[0.01-0.1]$ and $\epsilon_{\rm
B}=[10^{-4}-10^{-2}]$, which are consistent with the typical values derived in previous studies (e.g.,
Panaitescu \& Kumar 2002; Yost et al. 2003). The Compton parameter is assigned to a typical value $Y
=1$. The $E_{\rm K, iso}$ of GRBs are reported in Table 3.

With the derived $E_{\rm \gamma,iso}$ and $E_{\rm K,iso}$, one can define the total isotropic GRB
energy
\begin{equation}
\label{total_energy} E_{\rm GRB,iso} = E_{\rm \gamma,iso} + E_{\rm K,iso}.
\end{equation}

To study the true energetics of the GRBs, the jet collimation angle $\theta_j$ needs to be derived. We
derive this parameter using the time when a steepening break known as the ``jet break''  is observed in
the afterglow light curve. If such a break is not observed, we use the last observational time as the
lower limit of the jet-break time. The jet angle information was searched from the literature before
(Liang et al. 2008; Lu et al. 2012; the references in Table 1), which is adopted in our analysis. The
jet opening angle is derived by using (Frail et al. 2001; Zhang 2018)
\begin{eqnarray}
\label{theta_j} \theta_j  &= &  0.063 \left(\frac{t_j}{1 ~\rm day}\right)^{3/8}
\left(\frac{1+z}{2}\right)^{-3/8}\left(\frac{\eta_\gamma}{0.2}\right)^{1/8} \nonumber  \\& \times &
\left(\frac{E_{\rm
\gamma,iso}}{10^{53}~\rm erg}\right)^{-1/8} \left(\frac{n}{0.1 ~\rm cm^{-3}}\right)^{1/8},
\nonumber \\
\end{eqnarray}
where $t_j$ is the jet-break time (for non-detections, the last observational time is adopted to infer
the lower limit of the jet opening angle), and $\eta_\gamma$ is an efficiency conversion factor
($\eta_\gamma\equiv E_{\rm \gamma, iso}/E_{\rm K, iso}$). With the derived $\theta_j$, we correct the
isotropic values of various forms of energy by multiplying the values of the beaming fraction (Frail et
al. 2001)
\begin{eqnarray}
f_b = 1-\cos \theta_j \simeq (1/2) \theta_j^2,
\end{eqnarray}
so that $E_\gamma=E_{\rm \gamma,iso} f_b$, and $E_{\rm K} = E_{\rm K,iso} f_b$. We denote $E_{\rm GRB}$
as the total energy of a GRB, which is defined as
\begin{equation}
E_{\rm GRB}=E_{\rm \gamma}+E_{\rm K}= E_{\rm GRB, iso} f_{b},
\end{equation}
These are reported in Table 3.

\begin{figure}
\includegraphics[angle=0,scale=0.4]{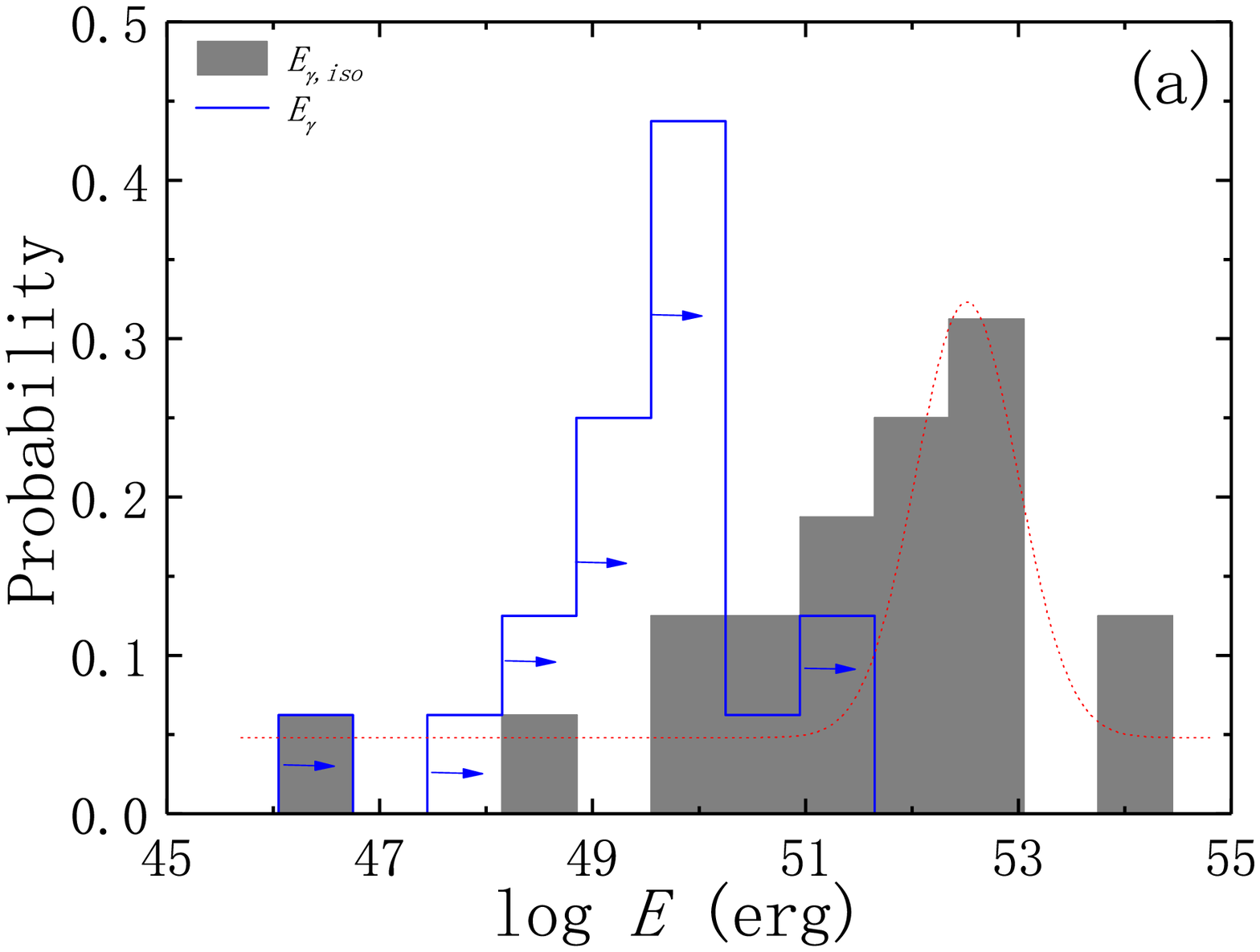}
\includegraphics[angle=0,scale=0.4]{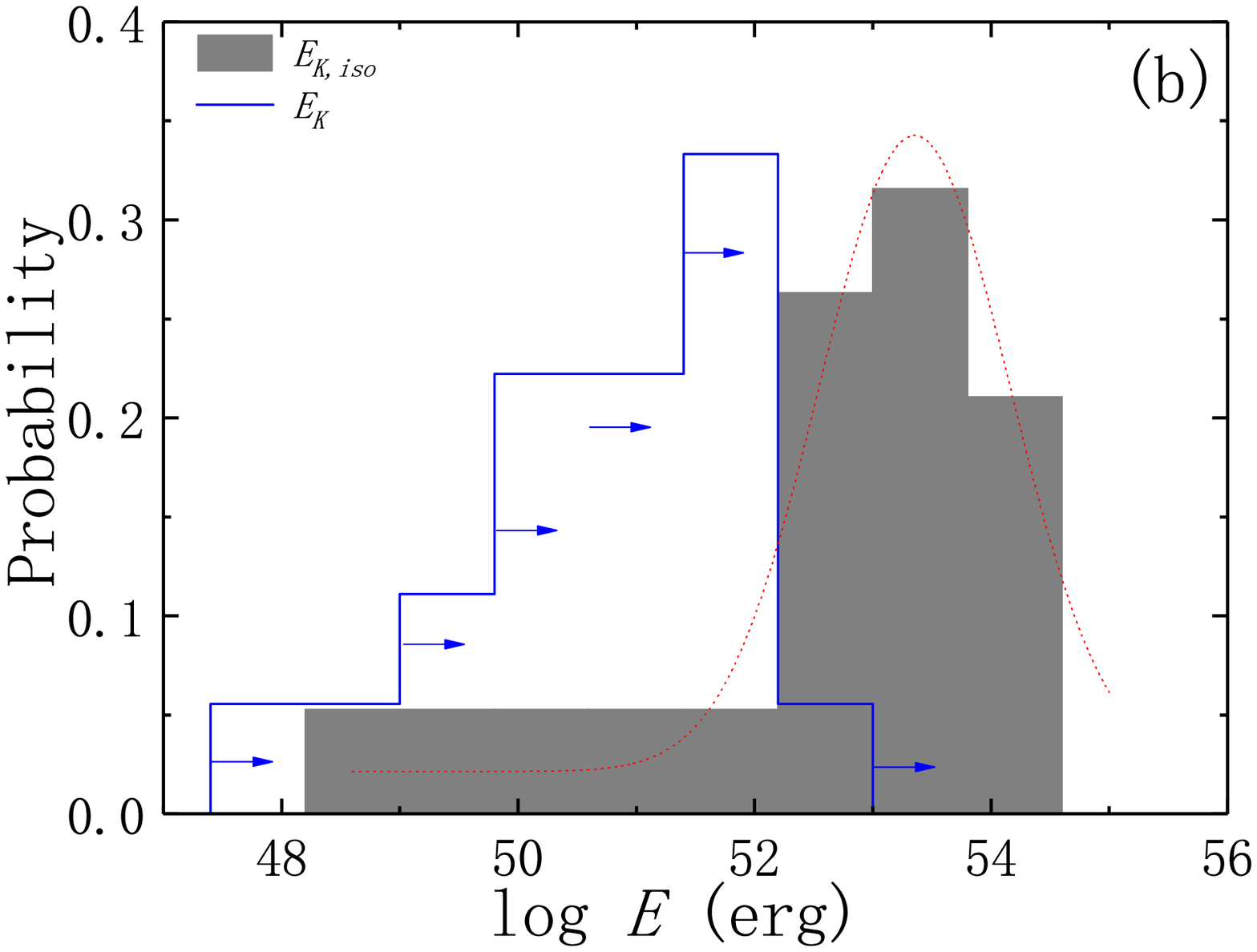}
\center\caption{Distributions of the $\gamma$-ray energy (a) and kinetic energy (b) for the isotropic
(gray-filled) and beaming-corrected values (blue solid line),
respectively. Best-fit Gaussian profiles are denoted by red dotted curves. The arrows are ower limits on the
energies after beaming correction.}
\label{fig:Energydis}
\end{figure}

The $E_{\rm \gamma, iso}$, $E_{\rm p}$, $E_{\rm K, iso}$ and $\theta_{\rm j}$ (or lower limit) of the
GRBs in our sample are summarized in Table 1. Figure \ref{fig:Energydis} presents the distributions of
the $\gamma$-ray energy and kinetic energy for the isotropic and beaming-corrected values,
respectively. With Gaussian fitting, the mean values of the isotropic energies are derived as $\rm log~
E_{\gamma, iso}=52.52\pm0.13$ erg and $\rm log~E_{K, iso}=53.35\pm0.13$ erg, respectively. Due to the
lack of jet-break detections in some GRBs of our sample, we could only plot the distributions of the
beaming-corrected $\gamma$-ray and kinetic energies using some of the lower limits, so that no reliable
Gaussian distributions can be derived.

\subsection{Physical parameters of SNe}

When identified, the peak luminosity and peak time of a SN associated with a GRB can be directly
inferred from the data. The nickel mass, explosion energy, and ejecta mass of a SN can be estimated
from bolometric light curves and spectral properties of the SN. These parameters can provide important
clues to understand the progenitors of the SN.

The bolometric light curve data of a SNe are collected from the literature. We apply the analytical
model of Arnett (1982) to derive the nickel mass and the ejecta mass. According to this model, the
luminosity of SN as a function of time reads
\begin{eqnarray}
L(t)& = & M_{Ni}\times \exp(-t^{2}/\tau^{2})\nonumber \\ & \times & \{\epsilon_{Ni}\int^{t/\tau}_{0}
A(x)dx+\epsilon_{Co}\int^{t/\tau}_{0}(B(x)-A(x))dx
\}\nonumber \\
\end{eqnarray}
where
\begin{eqnarray}
\cases{ A(x)=2x \cdot \exp (-2xy+2xs+x^2), \cr B(x)=2x \cdot \exp (-2xy+x^2),\cr \rm
y=\frac{\tau}{2\tau_{Ni}}, \cr \rm
s=\frac{\tau(\tau_{Co}-\tau_{Ni})}{2\tau_{Co}\tau_{Ni}}, \cr},
\end{eqnarray}
with $\epsilon_{Ni}=3.9\times10^{10}\rm~erg~s^{-1}g^{-1}$, $\epsilon_{Co}=6.78\times10^{9}\rm~
erg~s^{-1}g^{-1}$ (Sutherland \& Wheeler 1984; Cappellaro et al. 1997), and the decay life time of
$^{56}\rm Ni$ and $^{56}\rm Co$ are $\tau_{\rm Ni}=8.88$ days and $\tau_{\rm Co}=111.3$ days
(Taubenberger et al. 2006), respectively. Here, $\tau$ is the effective diffusion time that is related
to the opacity ($\kappa$), the ejecta mass ($M_{\rm ej}$), as well as the photospheric velocity
($v_{\rm ph}$) that can be determined by the width of the bolometric light curve, which reads
\begin{eqnarray}
\tau \approx \left(\frac{\kappa \cdot M_{\rm ej}}{\beta c\cdot v_{\rm ph}} \right)^{1/2}.
\end{eqnarray}
where $\beta$ is a constant of integration (Arnett 1982), $c$ is the speed of light, and the opacity
has an assumed typical value $\kappa=0.07 \rm~cm^{2}g^{-1}$ (Chugai 2000). We collect the $v_{\rm ph}$
and $M_{\rm ej}$ values from the literature and derive $M_{\rm Ni}$ based on light curve fitting. The
kinetic energy of the ejecta is derived as
\begin{eqnarray}
E_{\rm SN}=\frac{1}{2}M_{\rm ej} v^2_{\rm ph},
\end{eqnarray}
where we have assumed that the explosion is spherically symmetric. If one assumes that the SN explosion
is asymmetric and it is brighter near the polar region (i.e., the GRB jet direction), then the true
kinetic energy may be smaller by a factor of a few (e.g., from 2 to 5, Mazzali et al. 2014). Notice
that there are five SNe (SNe 2001ke, 2002lt, 2005nc, 2013ez, and 2013fu) that do not have enough data
(their SN signature was inferred from the optical bump in the late afterglow light curve). The
parameters of those cases are taken from the literature. All together, the derived SN parameters are
summarized in Table 2 and 3.

\begin{figure}
\includegraphics[angle=0,scale=0.4]{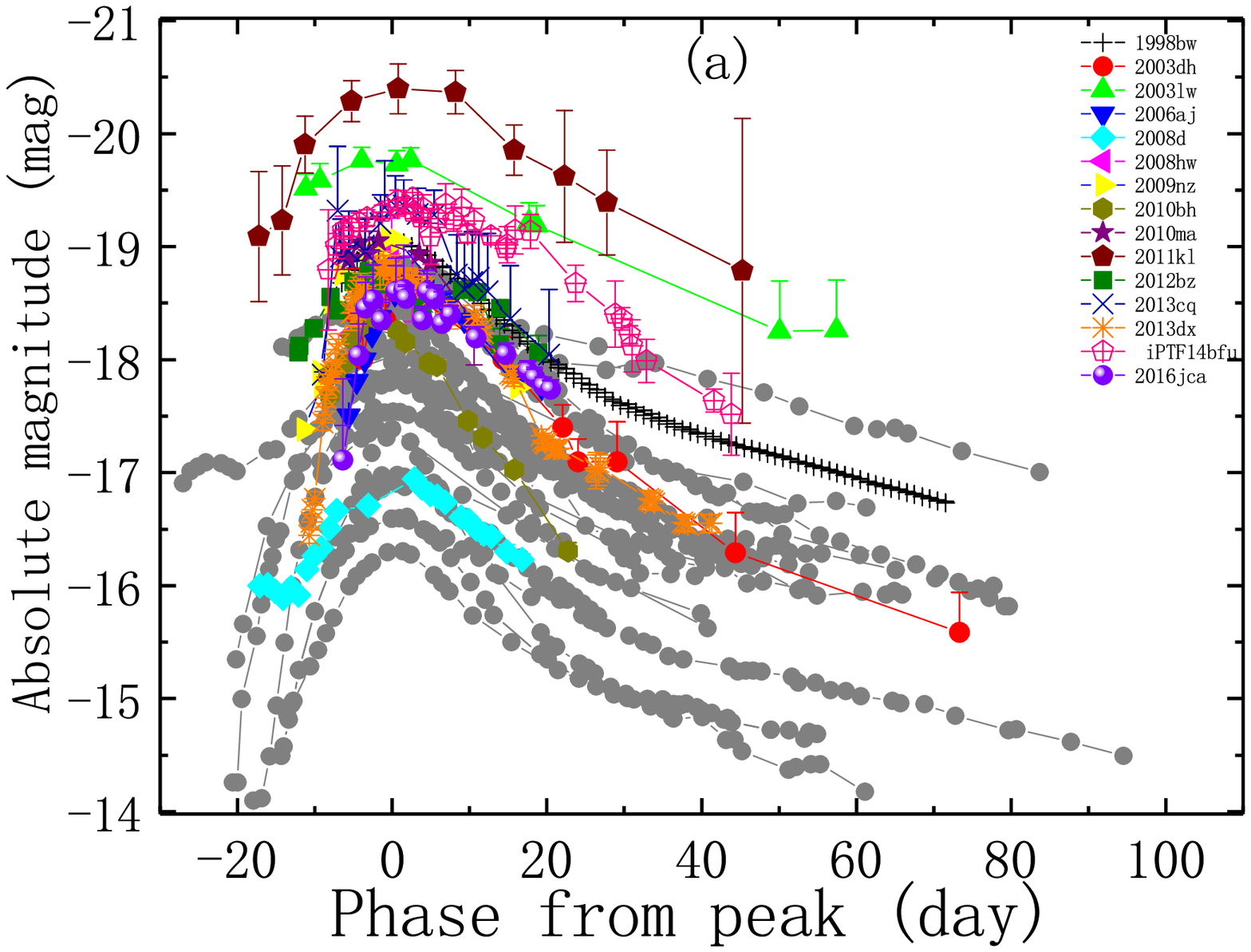}
\includegraphics[angle=0,scale=0.4]{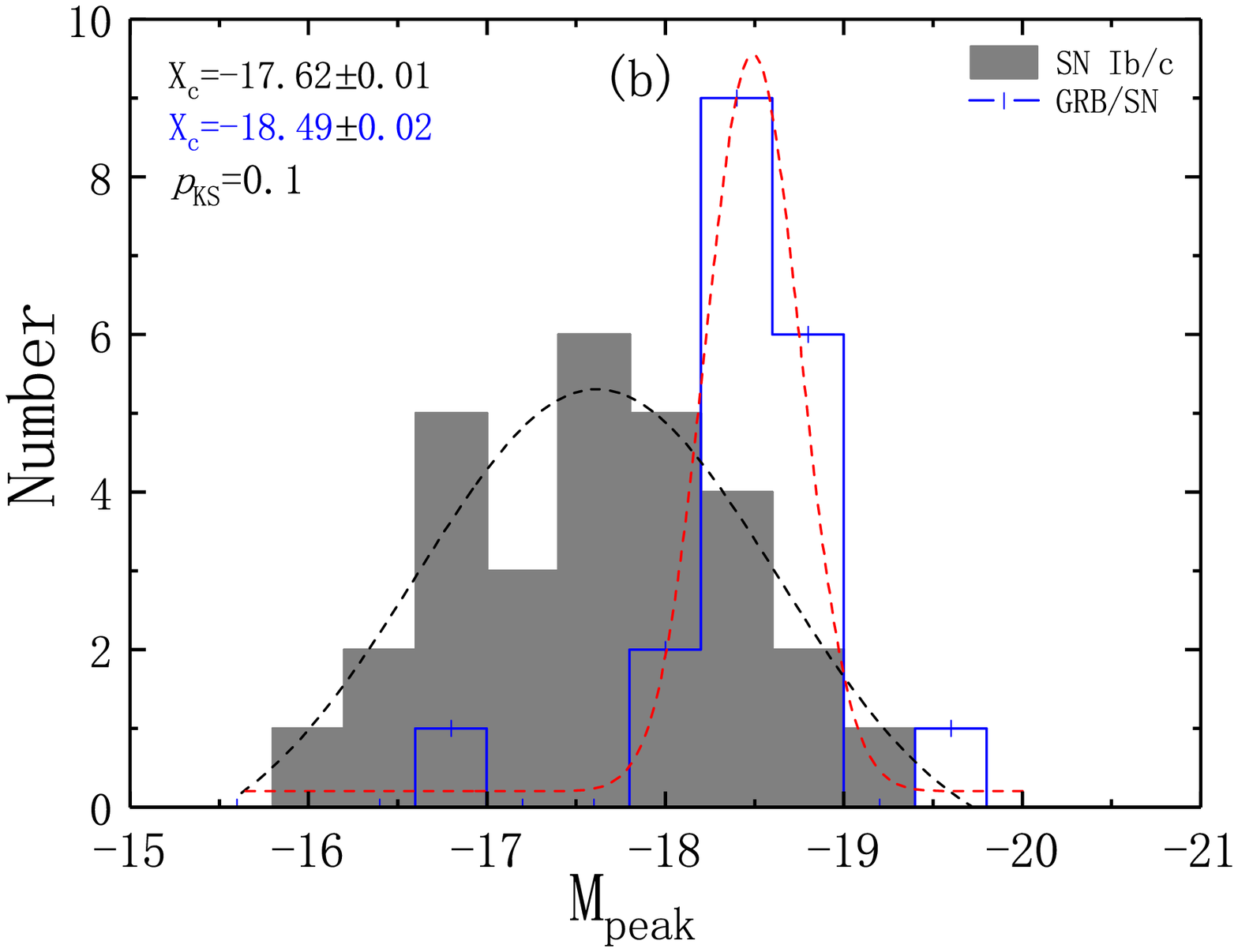}
\center\caption{Comparison of the bolometric light curves (a) and peak magnitudes (b) of the GRB-associated
SNe in our sample and other Type Ib/c SNe (gray). The data
of other Type Ib/c SNe are taken from Lyman et al. (2016). The dashed lines of (b) are the best Gaussian
fits.}
\label{fig:SNLC}
\end{figure}

Figure \ref{fig:SNLC} shows a comparison of the bolometric light curves (a) and peak magnitudes (b) of
the GRB-associated SNe in our sample and other Type Ib/c SNe (Lyman et al. 2016). The bolometric light
curves of the SNe (Figure \ref{fig:SNLC}(a)) are plotted with the zero time set at the peak time. For
comparison, we plot the bolometric light curves of other Type Ib/c SNe (taken from Lyman et al. 2016)
in gray. Figure \ref{fig:SNLC}(b) shows the distribution of the peak magnitude of GRB-associated SNe
(solid histogram) and that of other Type Ib/c (gray histogram). It can be seen that GRB-associated SNe
are systematically brighter than other Type Ib/c SNe. However, a K-S test shows that it is only
somewhat unlikely that the offset between the two distributions stems from random chance
($p_{K-S}=0.1$). Typically, $p$=0.01 is seen as reasonable, and only $p<$0.001 is seen as strong
evidence for two truly different distributions.

\subsection{Statistical correlations of the GRB/SN parameters}

To investigate possible relations between GRB parameters and SN parameters, we present a series of
scatter plots.

\begin{figure}
\includegraphics[angle=0,scale=0.4]{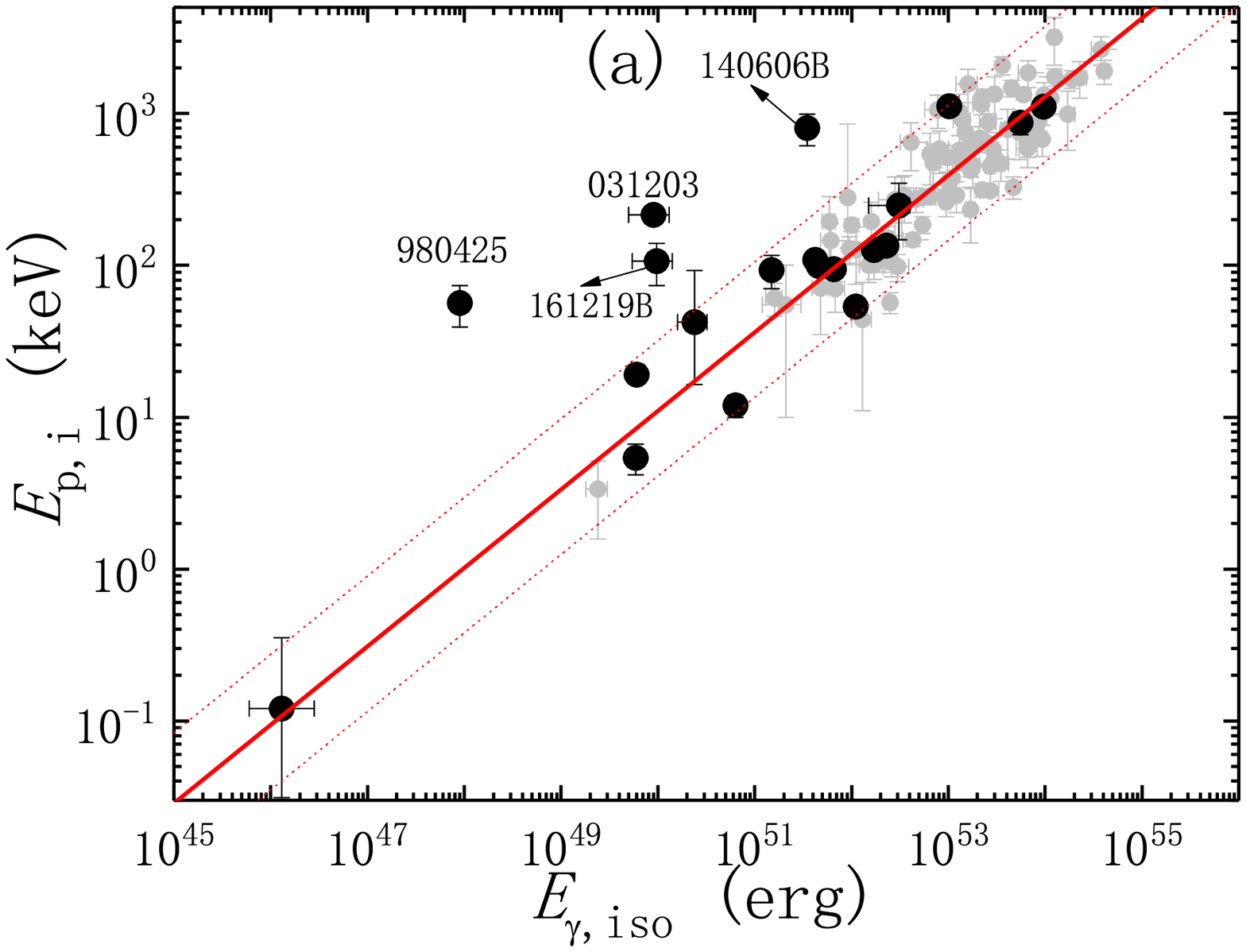}
\includegraphics[angle=0,scale=0.4]{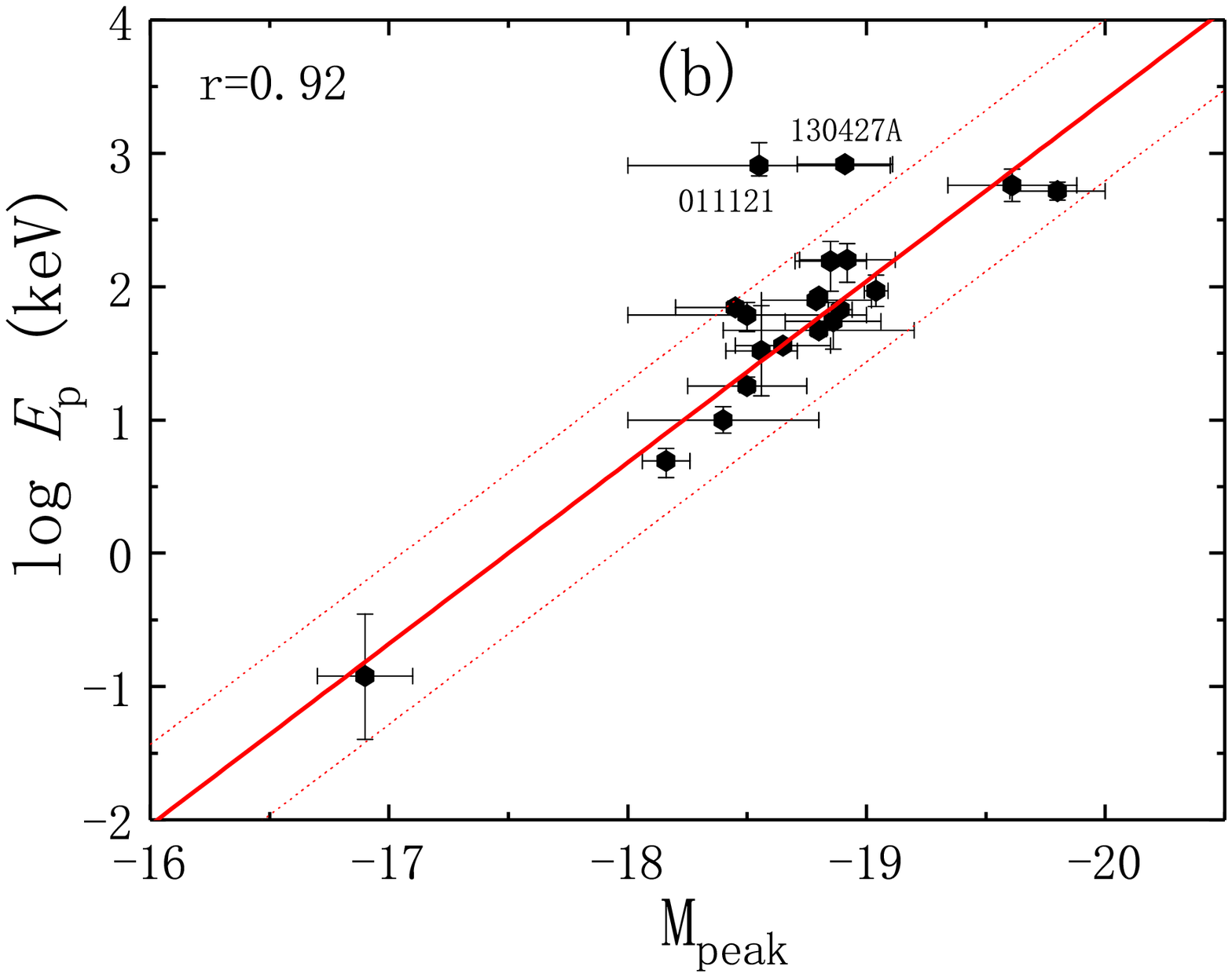}
\includegraphics[angle=0,scale=0.4]{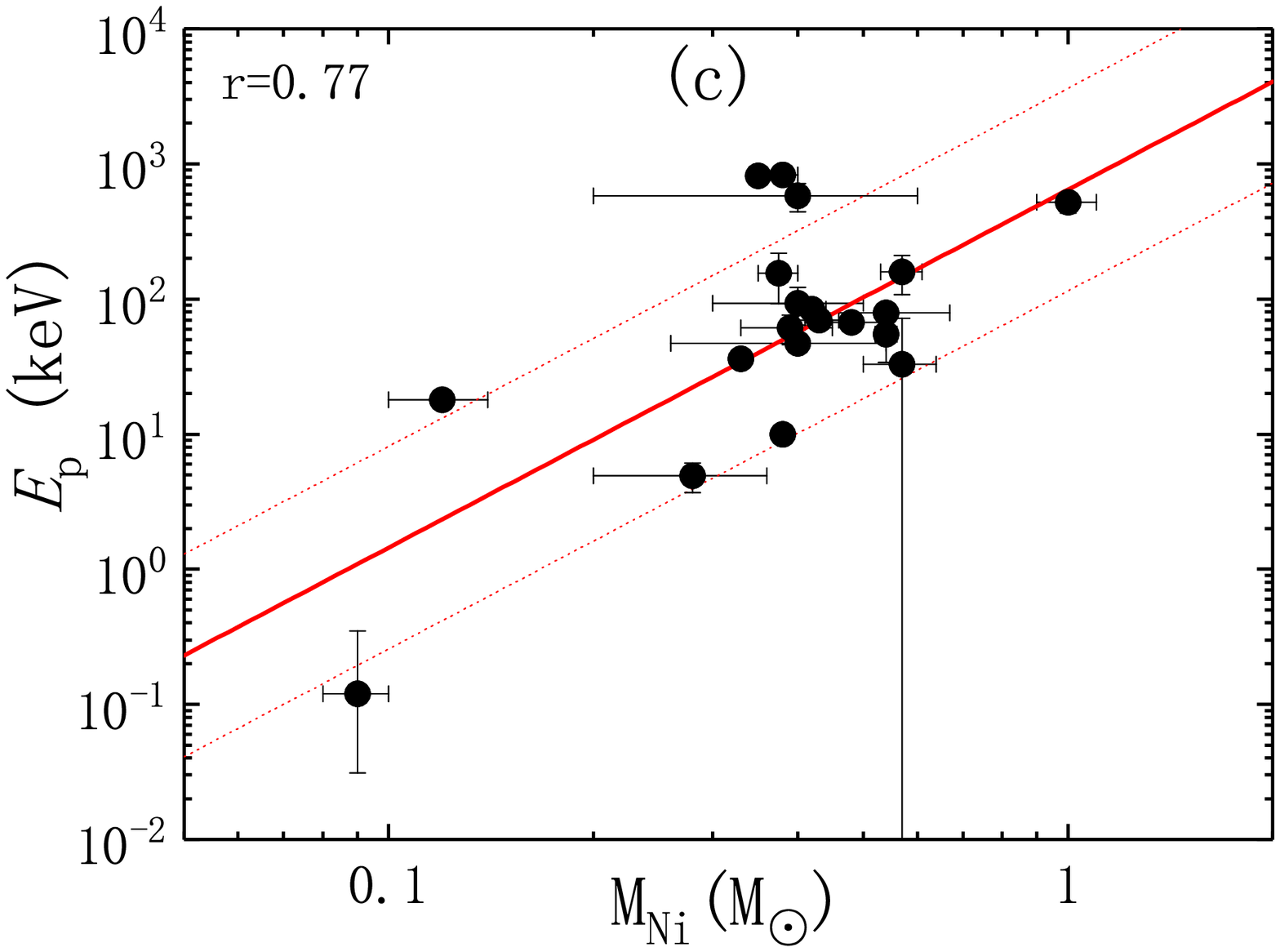}
\includegraphics[angle=0,scale=0.4]{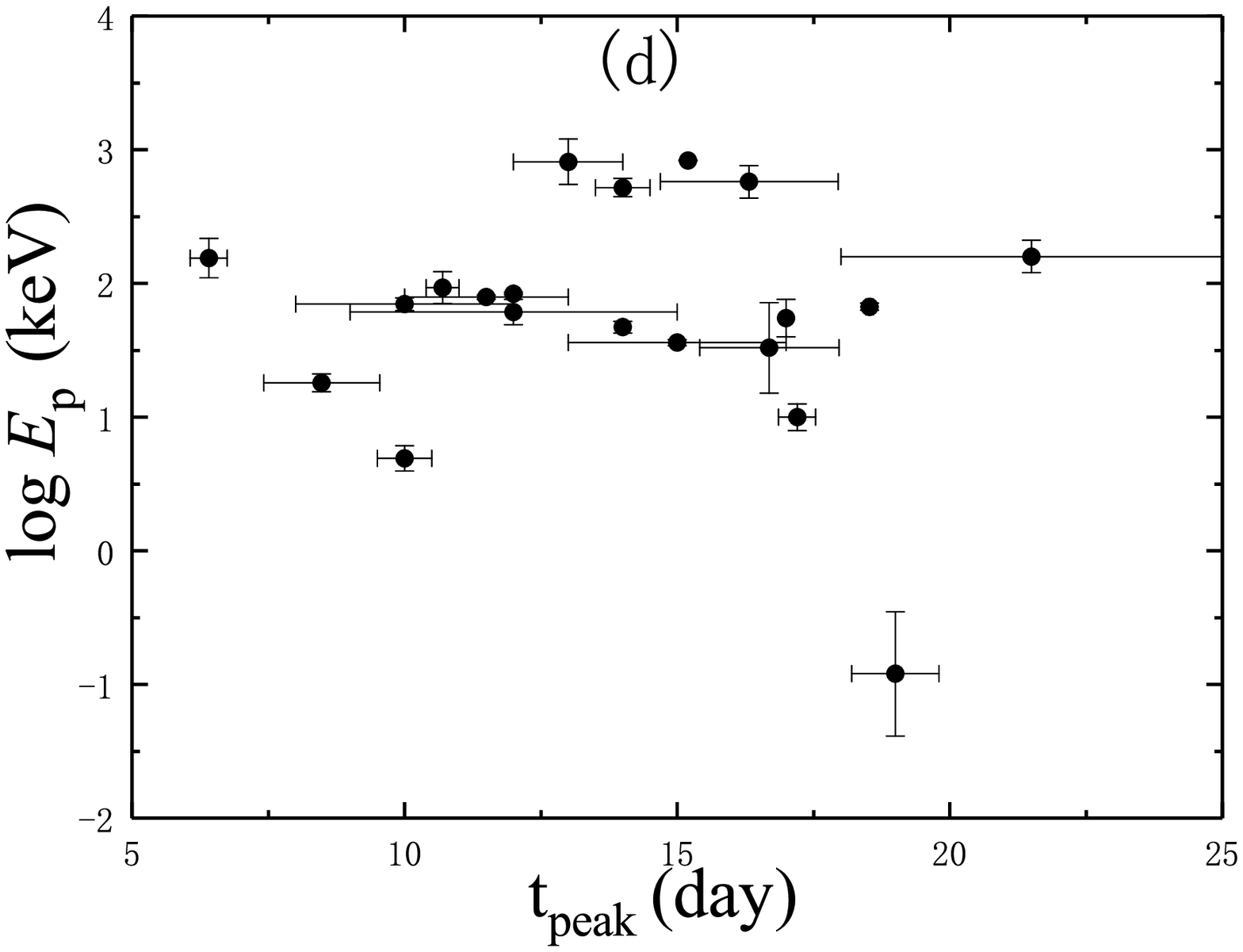}
\center\caption{Spectral peak energy ($E_{\rm p}$) of GRBs as a function of $E_{\rm \gamma, iso}$ (a),
$M_{\rm peak}$ (b), $t_{\rm peak}$ (c), and $M_{\rm Ni}$ (d).
When a correlation exists, a solid line is drawn for the best power-law fit (for panel (a) the outliers are
excluded). The dotted lines are the region of 2$\sigma$
deviation.}
\label{fig:Ep_correlation}
\end{figure}

We first investigate how the GRB spectral peak energy, $E_p$ is related to other parameters (e.g.,
$E_{p,i} - E_{iso}$, $E_p - M_{\rm peak}$, $E_p - t_{\rm peak}$, and $E_p - M_{\rm Ni}$). Figure
\ref{fig:Ep_correlation} (a) presents the well-known $E_{\rm \gamma, iso}-E_{\rm p,i}$ correlation
(i.e., the so-called Amati relation). Here, $E_{\rm p,i} = E_p (1+z)$ is the cosmological rest-frame
peak energy of the GRB. The data of typical long GRBs are taken from Amati et al. (2002) and Zhang et
al. (2009). Most GRBs in our sample fall into the 2$\sigma$ deviation region of the best-fit power-law
model for typical long GRBs, and some outliers (including the low-luminosity GRBs 980425, 031203,
140606B, and 161219B and the ultra-long GRB 111209A) are identified. These deviations may be intrinsic,
but it is also possible that they are due to an observational bias caused by the lack of detection of
soft X-ray emission associated with these GRBs (see e.g., Martone et al. 2017).

Next, we investigate the relation between $E_p$ and the supernova peak bolometric magnitude ($M_{\rm
peak}$). Li (2006) discovered a correlation between the two parameters using four pairs of GRB/SN
associations and found $E_{p} \propto M_{\rm peak}^{-1.987}$. We investigate this correlation using our
much-expanded sample and find that the correlation still exists, even though the slope is somewhat
shallower than the one found in Li (2006). The data and the best-fit correlation are shown in Figure
\ref{fig:Ep_correlation} (b). Our best-fit correlation gives
\begin{eqnarray}
\log~E_{p}= (-1.36\pm 0.14)M_{\rm peak}-(23.82\pm 2.53),
\end{eqnarray}
with the Pearson linear correlation coefficient $r=0.92$, corresponding to a probability $P=0.06$ for
zero correlation. This indicates that the $M_{\rm peak}$ and $ E_{\rm p}$ are strongly correlated. Cano
(2014) suggested a correlation between the brightness and width of the light curves of SNe associated
with GRBs as a Pearson's correlation coefficient is $r\sim 0.93$, which may be used as a standardizable
candles. As the $E_{\rm p} - M_{\rm peak}$ correlation has a similar $r$ value with that of Cano
(2014), it may be used as a potential standard candle as well.

In Figure \ref{fig:Ep_correlation} (c), we plot $E_{\rm p}$ against $M_{\rm Ni}$, the mass of $^{56}\rm
Ni$. We find that $E_{\rm p}$ is also correlated with $M_{\rm Ni}$ with a large systematic error, i.e.,
\begin{eqnarray}
\rm log~E_{p}= (2.65\pm 0.65)log~M_{Ni}+(2.81\pm 0.27),
\end{eqnarray}
with a Pearson's linear correlation coefficient $r=0.77$, corresponding to a probability $P=0.23$ for
zero correlation. In Figure \ref{fig:Ep_correlation} (d), we plot $E_p$ against $t_{\rm peak}$. No
significant trend is found.

\begin{figure}
\includegraphics[angle=0,scale=0.4]{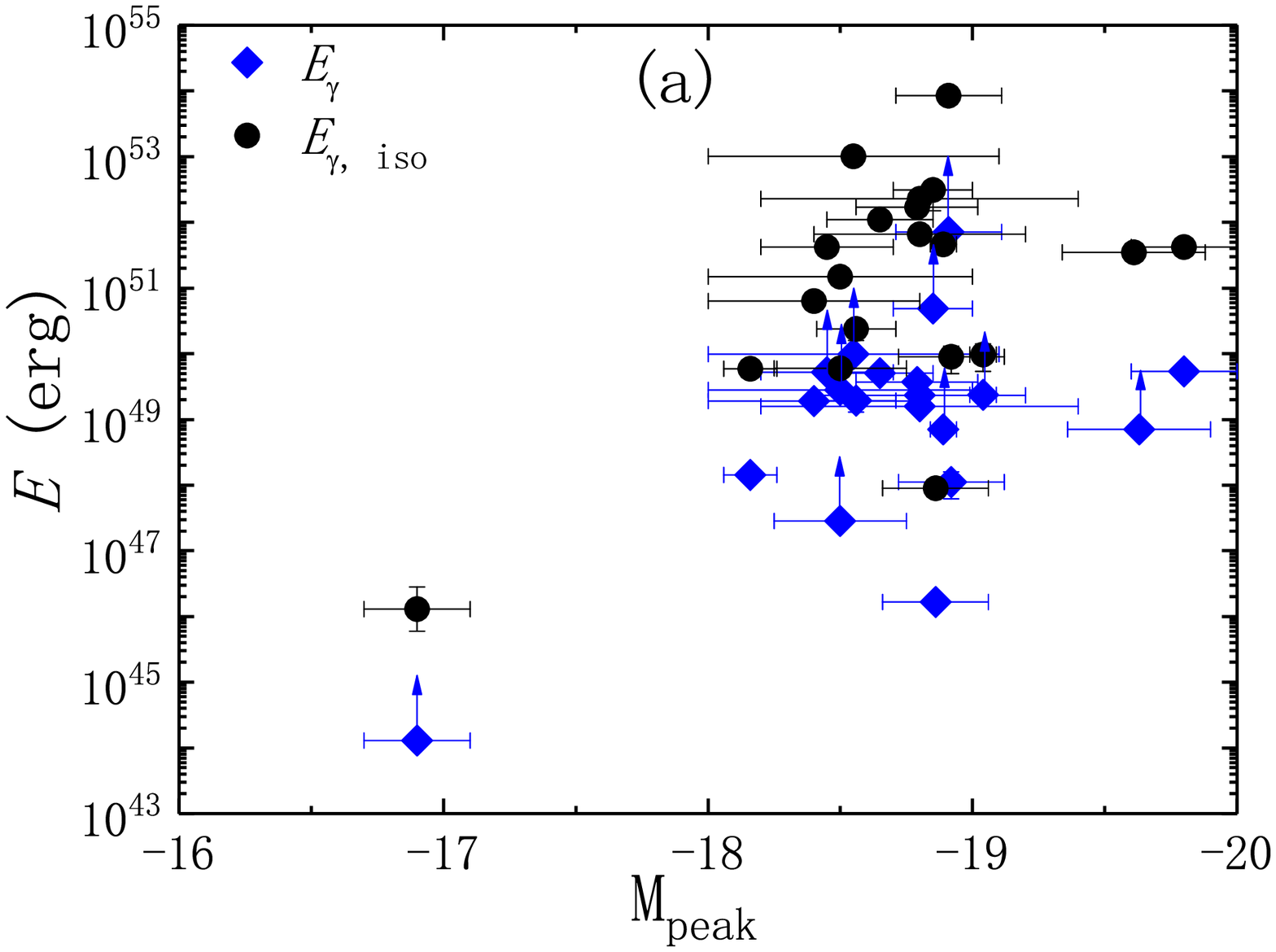}
\includegraphics[angle=0,scale=0.4]{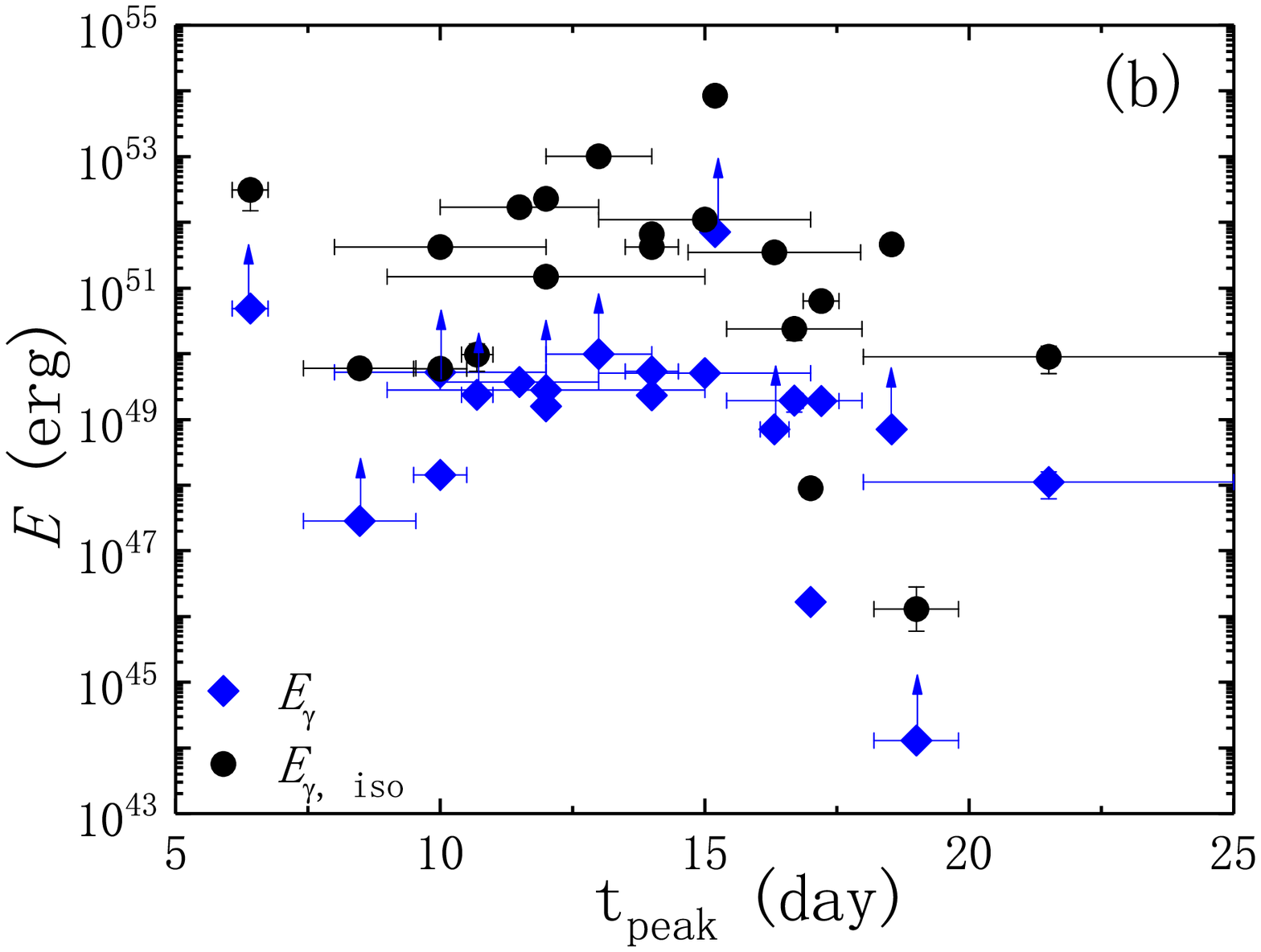}
\includegraphics[angle=0,scale=0.4]{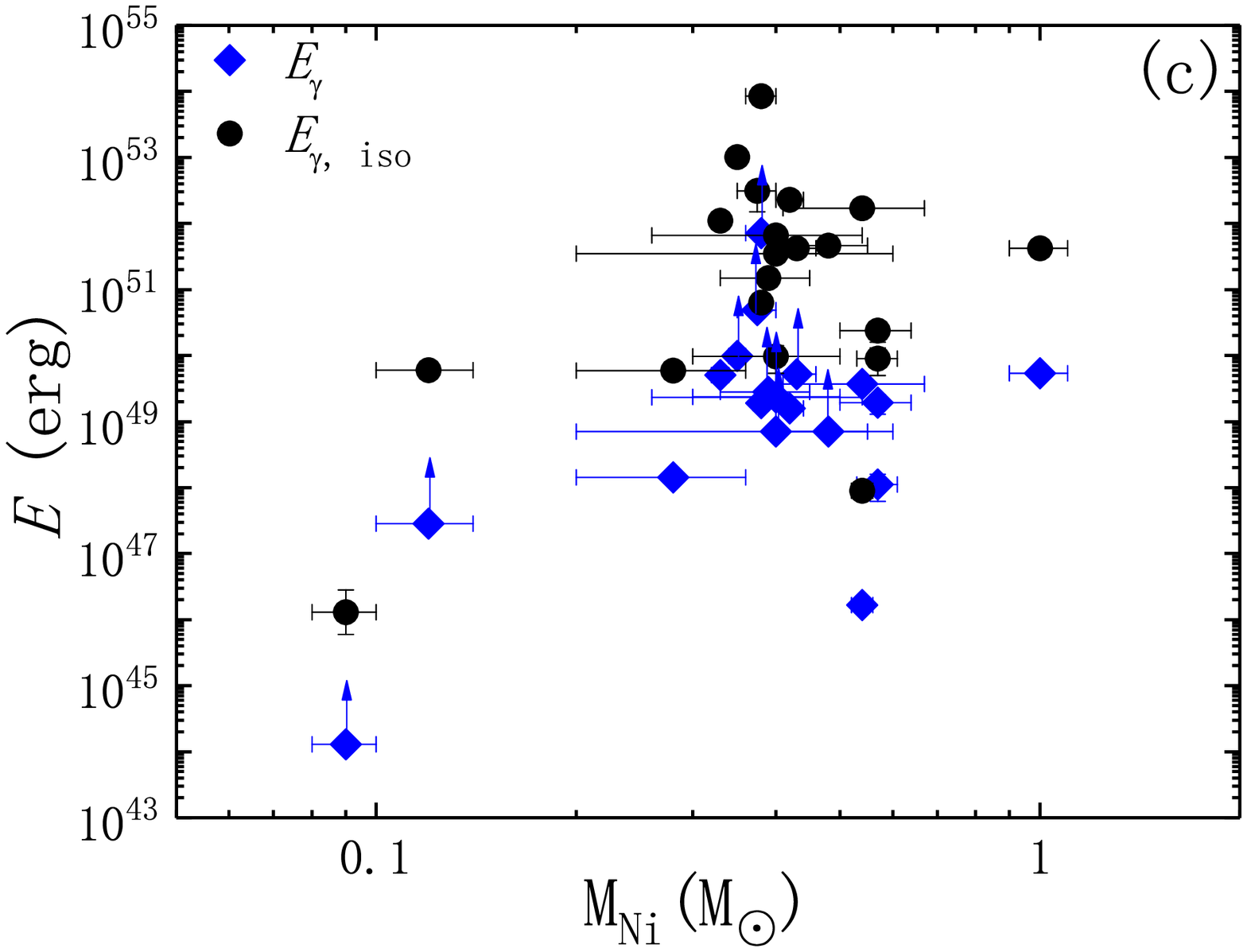}
\center\caption{Isotropic (black dots) and beaming-corrected (blue diamonds) prompt $\gamma$-ray
emission
energies vs. $M_{\rm peak}$ (a), $t_{\rm peak}$ (b), and
$M_{\rm Ni}$ (c). The blue arrows denote the lower limits of $E_{\rm \gamma}$.}
\label{fig:Eiso_correlation}
\end{figure}

\begin{figure}
\includegraphics[angle=0,scale=0.4]{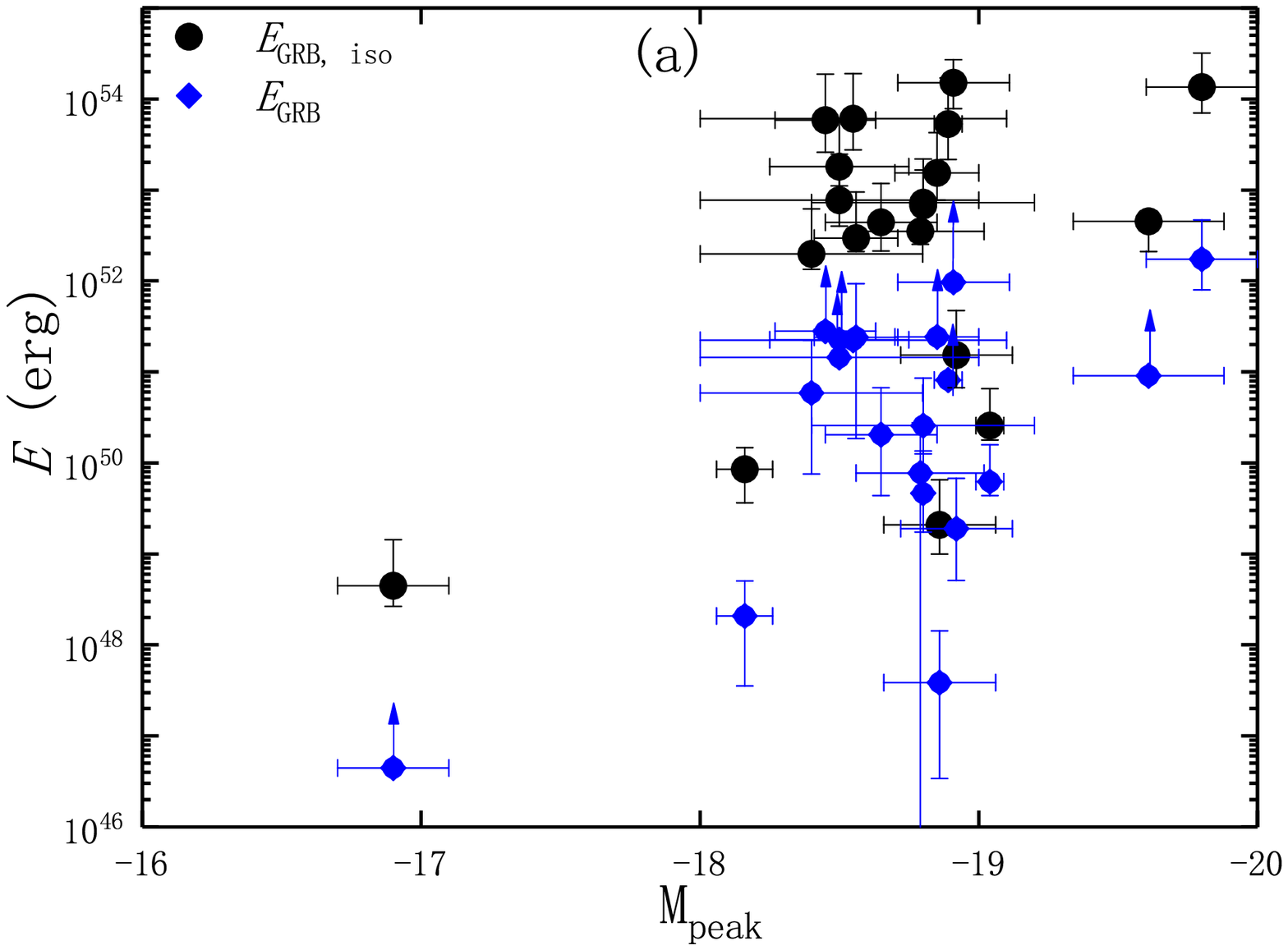}
\includegraphics[angle=0,scale=0.4]{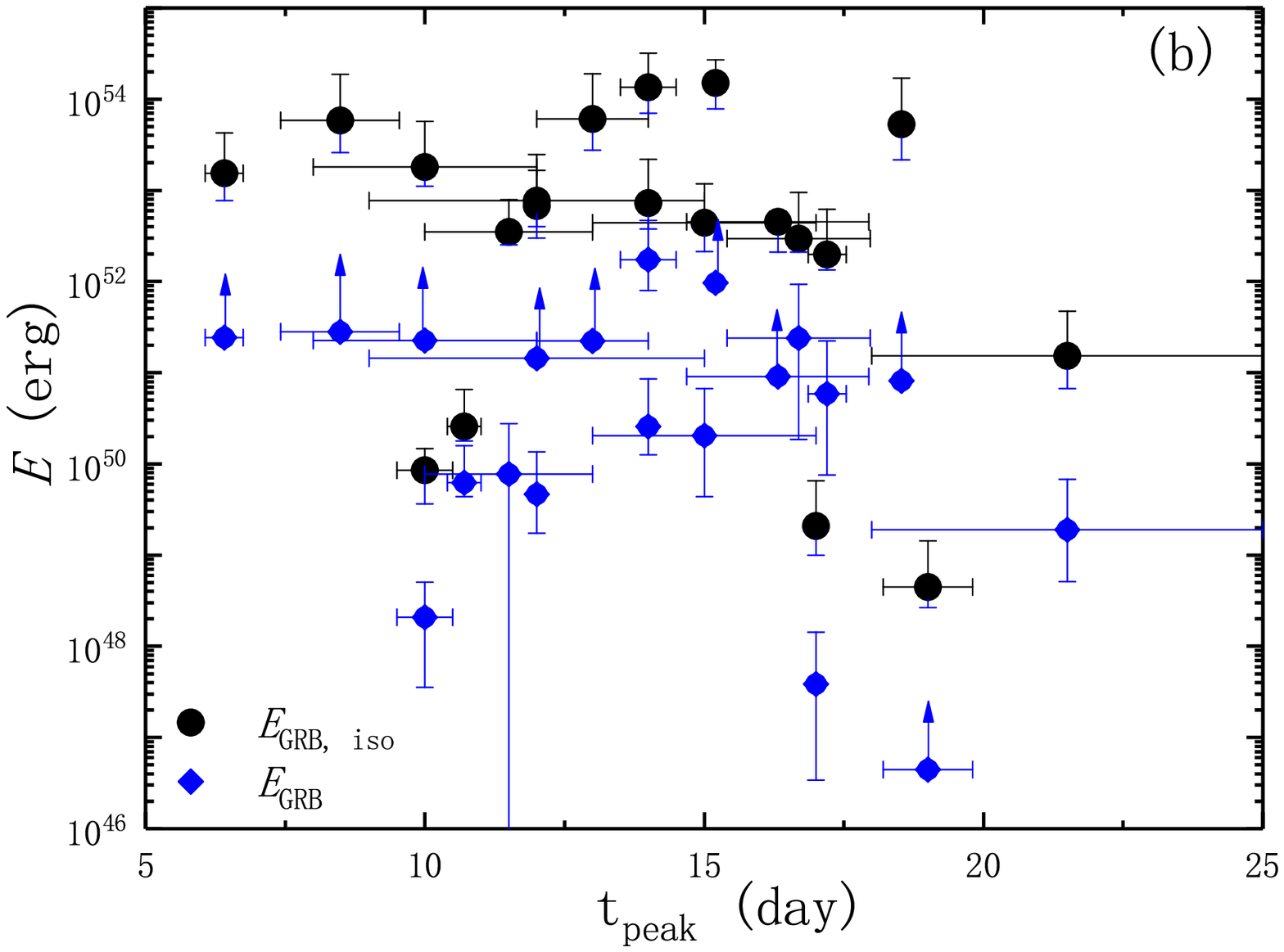}
\includegraphics[angle=0,scale=0.4]{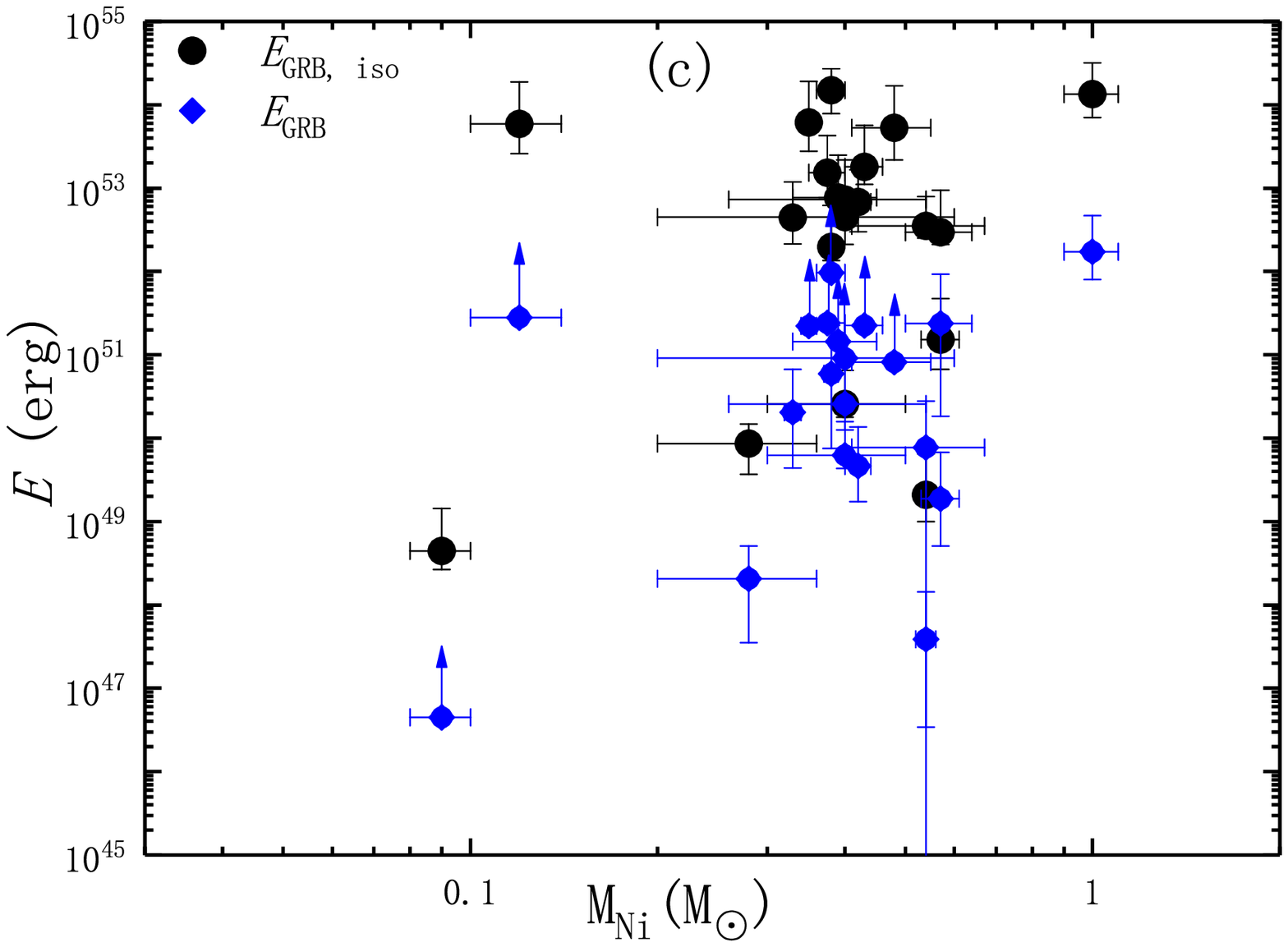}
\center\caption{Similar to Figure \ref{fig:Eiso_correlation}, but for the total GRB energies.}
\label{fig:EGRB_correlation}
\end{figure}

As $E_{\rm \gamma,iso}$ is strongly related to $E_p$ (Amati relation, Fig.\ref{fig:Ep_correlation}
(a)), we plot $E_{\rm \gamma,iso}$ (or beaming-corrected $\gamma$-ray energy $E_{\gamma}$) against
$M_{\rm peak}$, $t_{\rm peak}$, and $M_{\rm Ni}$ in Fig. \ref{fig:Eiso_correlation}. No significant
correlations between them are observed. In Fig. \ref{fig:EGRB_correlation}, we also plot the the total
GRB energy ($E_{\rm GRB, iso}$ and $E_{\rm GRB}$) against the three parameters. Again, no significant
correlations are observed.

\begin{figure}
\includegraphics[angle=0,scale=0.4]{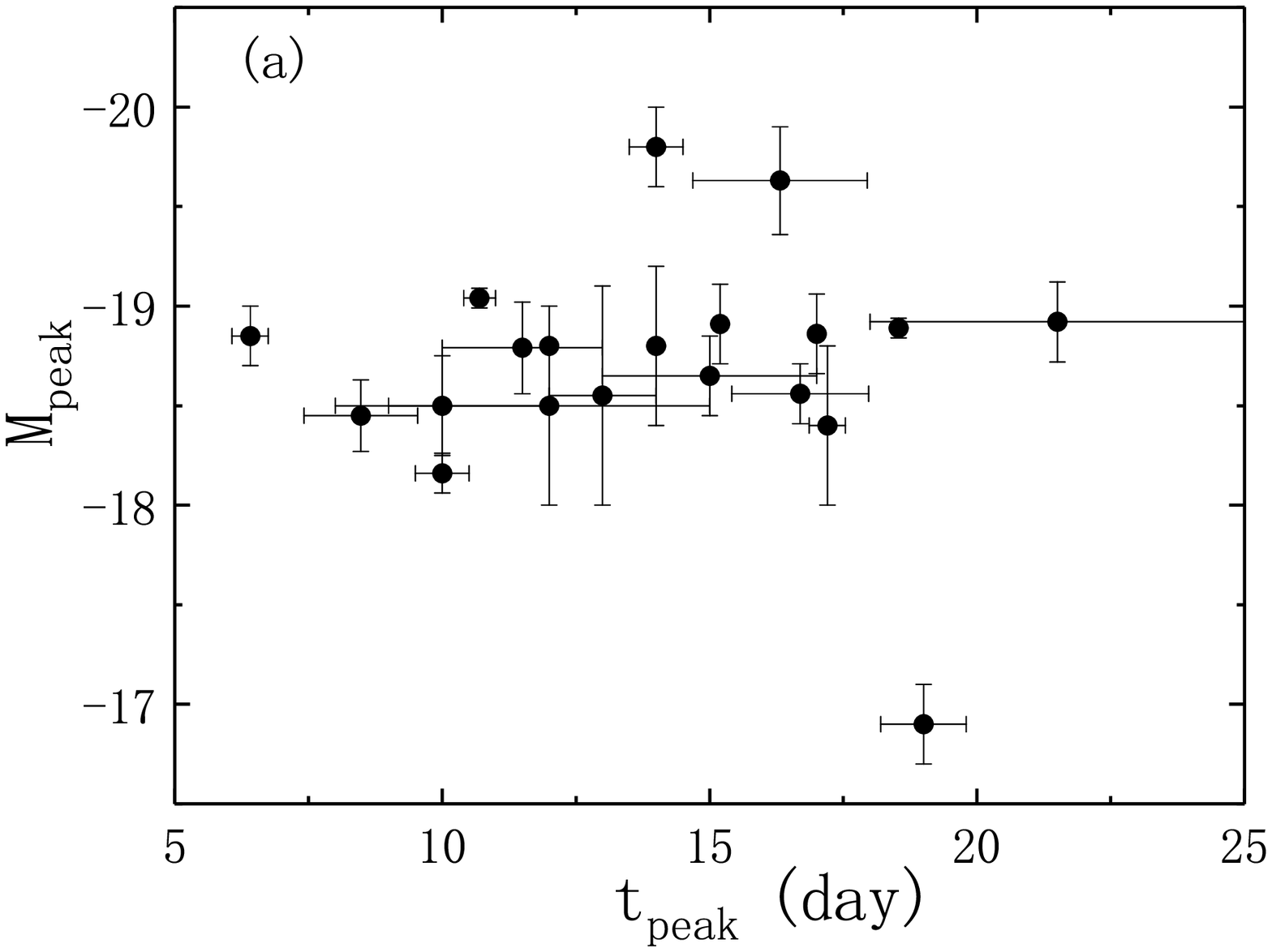}
\includegraphics[angle=0,scale=0.4]{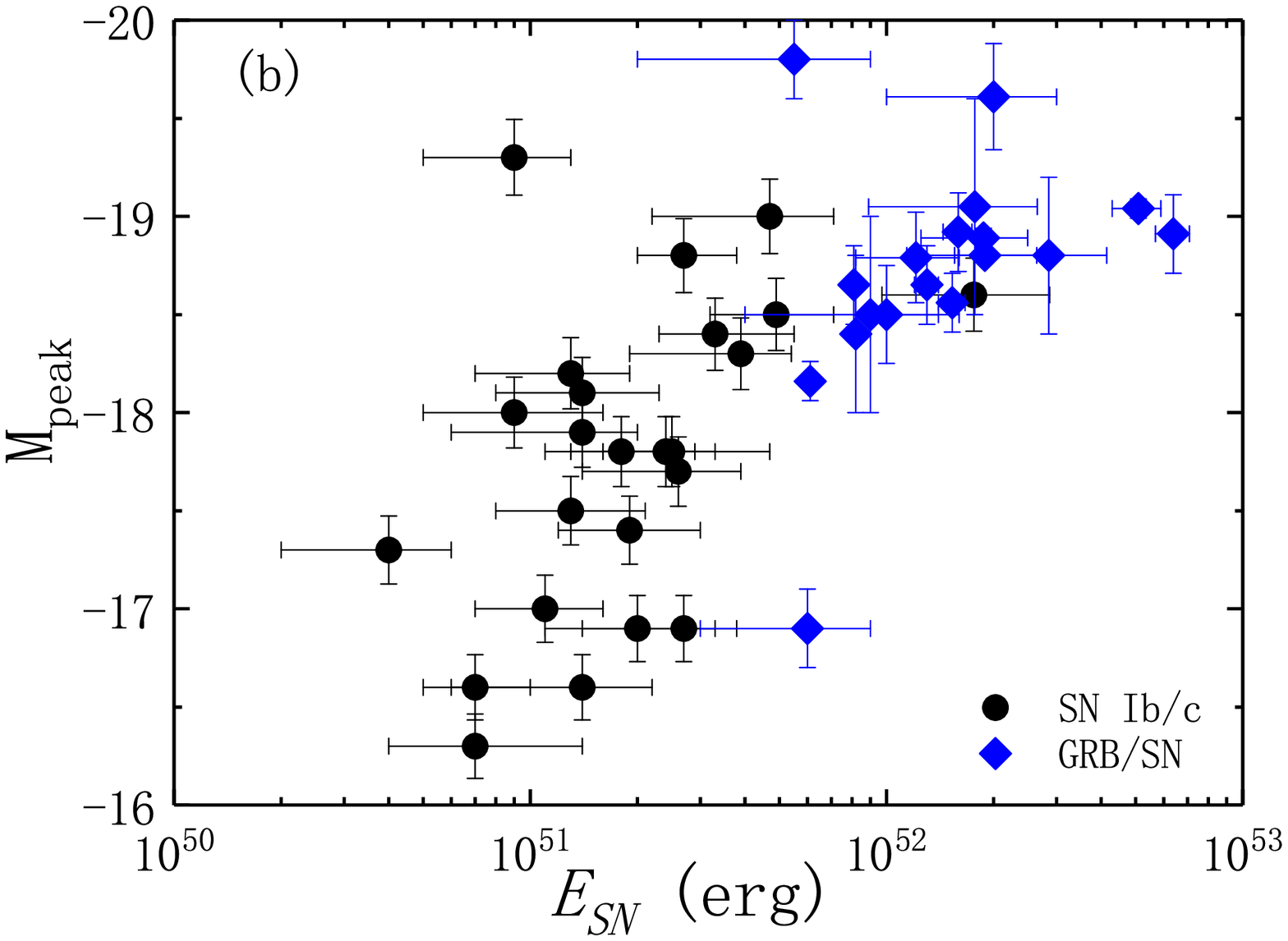}
\includegraphics[angle=0,scale=0.4]{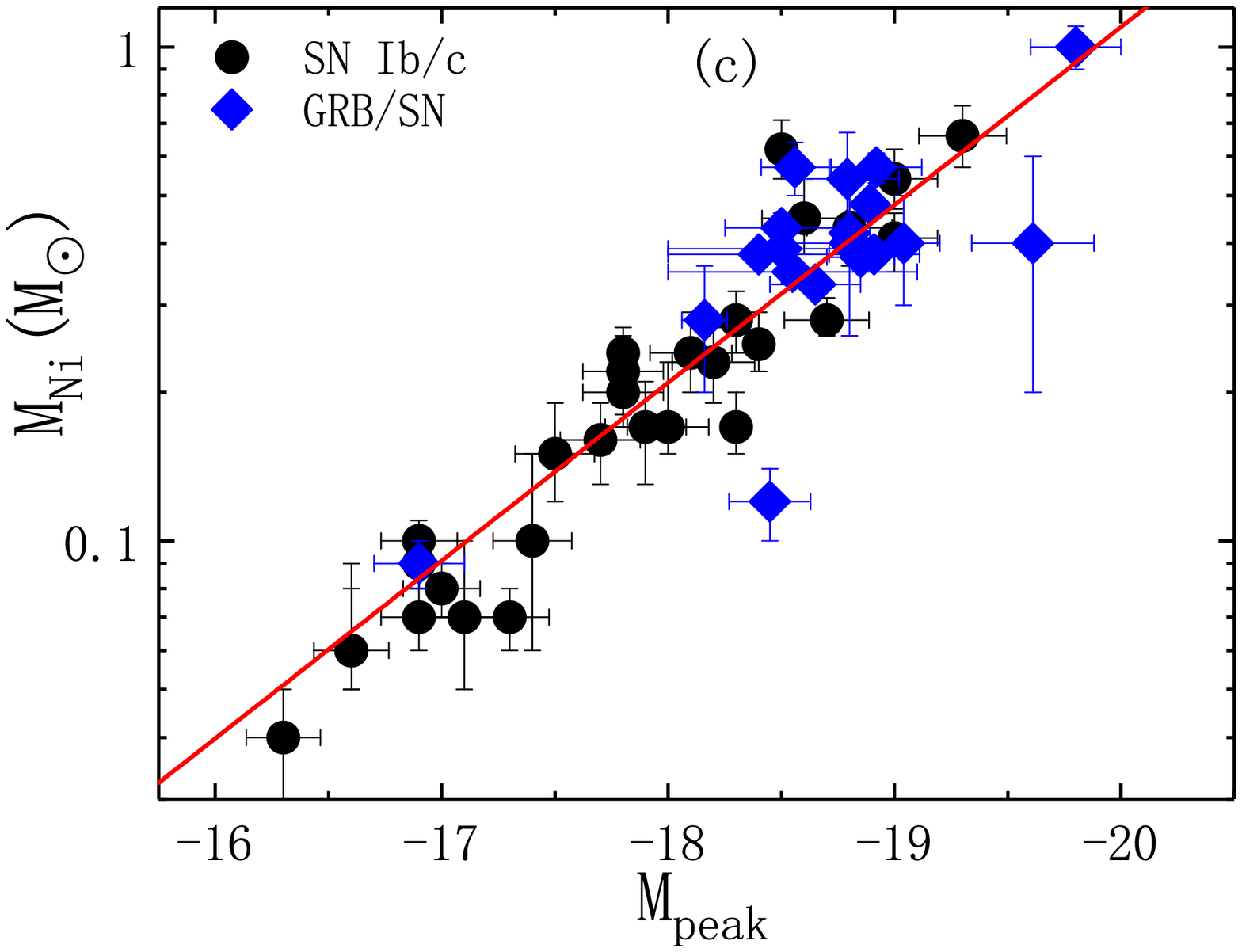}
\center\caption{Peak magnitude ($M_{\rm peak}$) of SNe as a function of $t_{\rm peak}$ (a), $E_{\rm SN}$
(b),
and $M_{\rm Ni}$ (c). The black dots and blue diamonds
denote our sample and other Type Ib/c SNe without GRB association, respectively. The solid red line is the
best power-law fit when an apparent correlation is seen.}
\label{fig:ESN_correlation}
\end{figure}

We also investigate some possible correlations among SN parameters. Figure \ref{fig:ESN_correlation}(a)
shows $M_{\rm peak}$ as a function of $t_{\rm peak}$ for SNe in our sample. No apparent correlation is
seen. Figure \ref{fig:ESN_correlation}(b) shows $M_{\rm peak}$ against $E_{\rm SN}$ for our sample and
other Type Ib/c SNe without associated GRBs. A rough trend of correlation is seen, but statistically no
significant correlation can be claimed. Figure \ref{fig:ESN_correlation}(c) shows a strong correlation
between $M_{\rm peak}$ and $M_{\rm Ni}$, which reads
\begin{eqnarray}
\rm log~M_{\rm Ni}= (-0.36\pm 0.14)M_{\rm peak}-(7.16\pm 0.61),
\end{eqnarray}
with the Pearson's linear correlation coefficient of $r=0.93$, corresponding to a probability $P=0.07$
for zero correlation. This is readily understandable, as $M_{\rm Ni}$ is derived making use of the peak
luminosity.

\section{Energy partition between GRB and SN, and the case of a magnetar central engine}

\subsection{Energy partition}

\begin{figure}
\includegraphics[angle=0,scale=0.6]{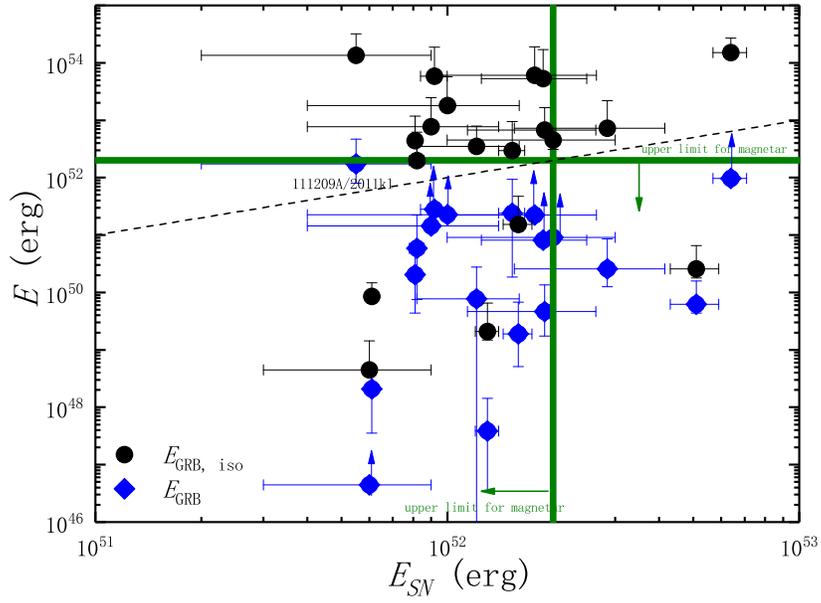}
\center\caption{$E_{\rm GRB,iso} / E_{\rm GRB}$ vs. $E_{\rm SN}$ in our sample. The dashed line denotes the
equality line. The vertical and horizontal lines are the
upper limit of the magnetar energy budget.}
\label{fig:E_GRBSN}
\end{figure}

Figure \ref{fig:E_GRBSN} shows $E_{\rm GRB,iso}$ vs. $E_{\rm SN}$ and $E_{\rm GRB}$ vs. $E_{\rm SN}$ in
our sample. The dashed line denotes the equality line. One can see that without beaming correction,
$E_{\rm GRB,iso}$ has a wide spread of more than three orders of magnitude. After beaming correction,
the distribution of $E_{\rm GRB}$ becomes narrower, now being within two orders of magnitude. There is
no direct correlation between the GRB energy and SN energy. In general, the SN energy is greater than
the GRB energy. Only the ultra-long GRB 111209A - super-luminous SN 2011kl association shows the
opposite trend, i.e., more energy is given to the GRB than the SN.

\begin{figure}
\includegraphics[angle=0,scale=0.6]{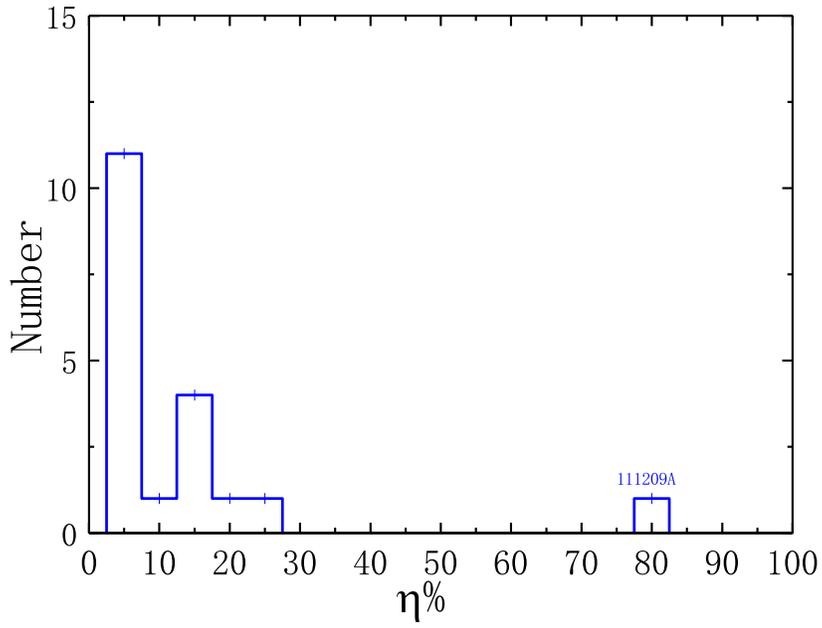}
\center\caption{Distribution of $\eta$ in our sample.}
\label{fig:FRACTION}
\end{figure}

One can also define the efficiency of GRB/SN events, i.e.,
\begin{eqnarray}
\eta=\frac{E_{\rm GRB}}{E_{\rm GRB}+E_{\rm SN}}
\end{eqnarray}
to denote the energy partition. Figure \ref{fig:FRACTION} shows the distribution of $\eta$ for our
sample. We find that the $\eta$ is usually less than 0.3 (with the center value $\sim 0.1$). The GRB
111209A/SN 2011kl system has $\eta$ greater than 70\% (see Table 3).

\subsection{The case of a magnetar central engine}

The remnant of massive-star core-collapse that produces a GRB is thought to be either a black hole or a
rapidly rotating magnetar. The SN is believed to be usually powered by the decay of $^{56}\rm Ni$
(Maeda \& Tominaga 2009). However, the existence of a magnetar as the central engine can inject
additional energy to power the SN, making it brighter (e.g., Bucciantini et al. 2009; Thompson et al.
2010; Woosley 2010; Metzger et al. 2011; Dessart et al. 2012; Wang et al. 2015, 2016).

\begin{figure}
\includegraphics[angle=0,scale=0.4]{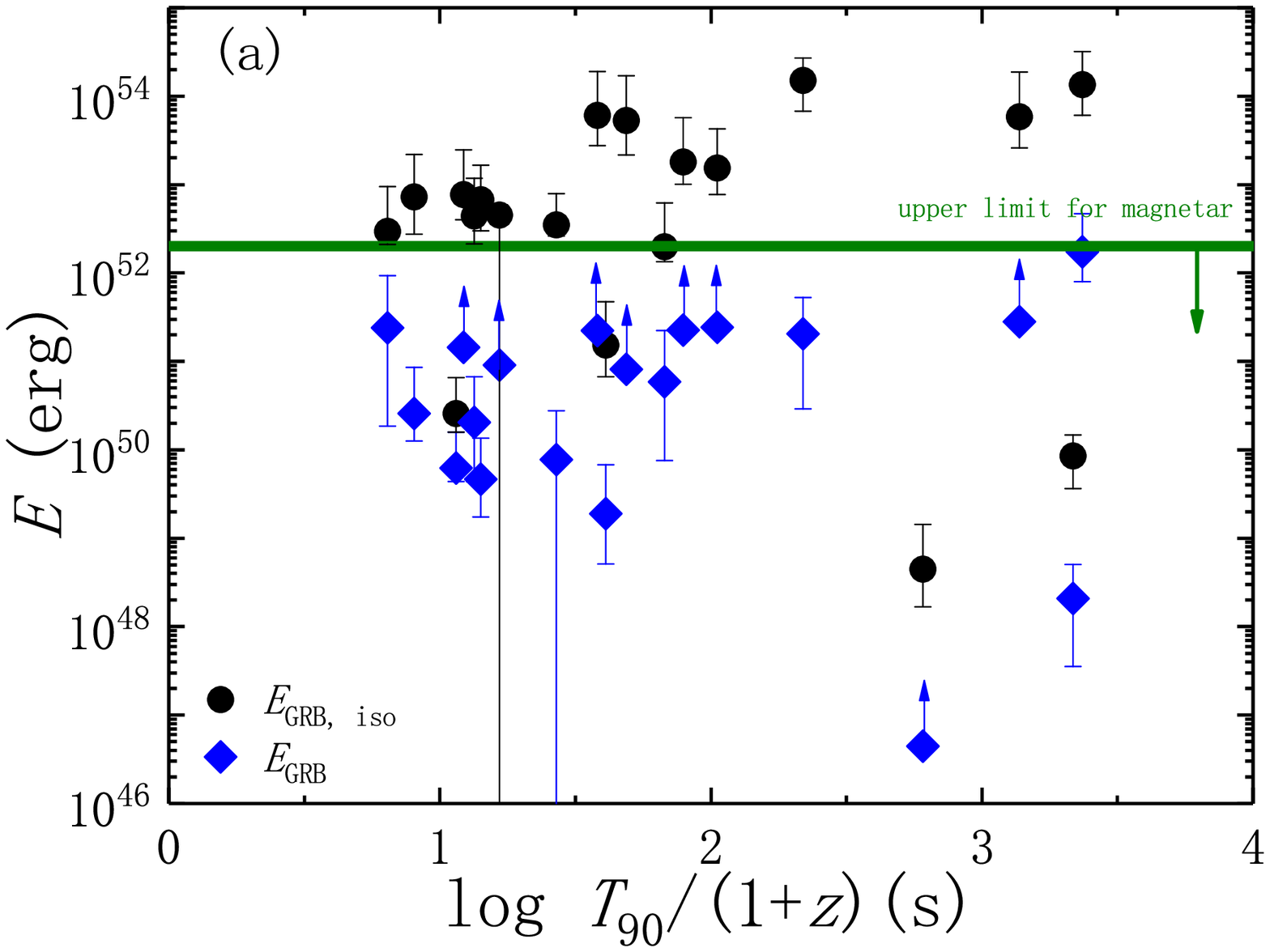}
\includegraphics[angle=0,scale=0.4]{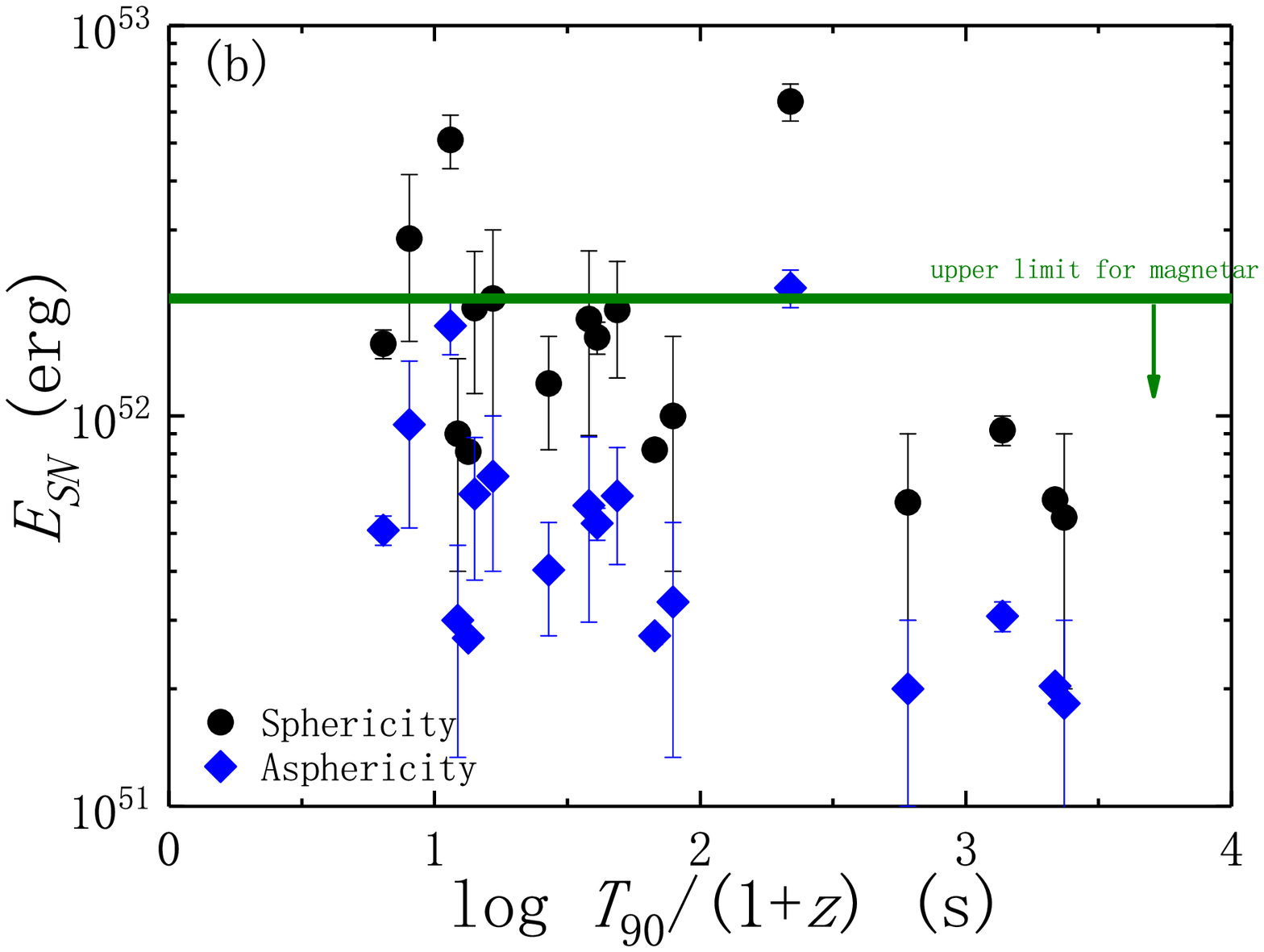}
\includegraphics[angle=0,scale=0.4]{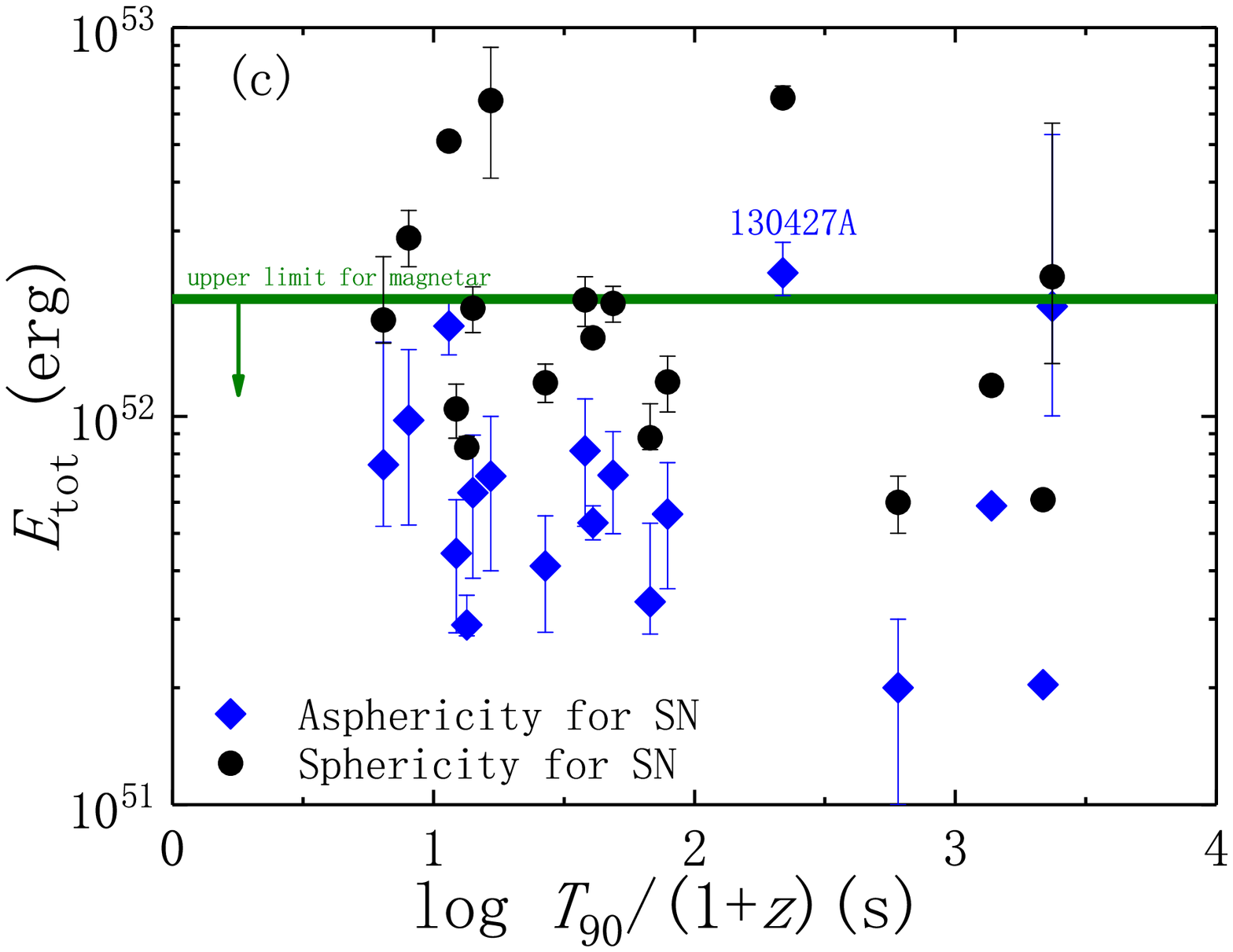}
\center\caption{$E_{\rm GRB, iso}/E_{\rm GRB}$, $E_{\rm SN}$ and $E_{\rm tot}$ against GRB rest-frame
duration. The horizontal line is the upper limit of the magnetar
energy budget.}
\label{fig:TOTALGRBSN}
\end{figure}

A magnetar engine has two specific predictions. First, due to significant energy injection, the SN is
expected to be brighter than a normal SN; the latter has a neutron-star engine with a much less spin
energy than a millisecond magnetar. This is supported by our data (see Fig.\ref{fig:SNLC}(b)). The
second prediction is that the total energy budget of the system (including both the GRB and the SN)
should not exceed the maximum spin energy of the millisecond magnetar, which is $\rm E_{rot}\sim
2\times 10^{52}$ erg for a magnetar with $M_{\rm NS}\sim1.4M_{\odot}$ and initial spin period $P_0\sim
1$ ms. To test this prediction, in Figure \ref{fig:TOTALGRBSN}, we plot the total isotropic GRB energy
($E_{\rm GRB,iso}$), total beaming-corrected GRB energy ($E_{\rm GRB}$), the SN energy ($E_{\rm SN}$),
as well as the total explosion energy of the GRB/SN system ($E_{\rm tot} = E_{\rm GRB}+E_{\rm SN}$) as
a function of the rest-frame GRB duration $T_{90}/(1+z)$. One can see that the majority of systems have
a total energy below the maximum energy budget of a millisecond magnetar. This can be also see in
Figure \ref{fig:E_GRBSN}, where the maximum energy budget lines are also plotted for $E_{\rm GRB}$ and
$E_{\rm SN}$, respectively.

Note that the SN energy  $E_{\rm SN}$ is calculated by assuming a spherical symmetry for the explosion.
We find that $E_{\rm SN}$ of most SNe are below or close to the maximum rotation energy of magnetar,
except in three cases (GRB 021211/SN 2002lt, GRB 130427A/SN 2013cq, and GRB 161219B/SN 2016jca) that
exceed this upper limit. In particular, the lack of a jet break 2.5 years after the burst trigger in
GRB 130427A (De Pasquale et al. 2016b) indeed poses a challenge to the total energy budget of this
event (a structured jet may somewhat alleviate this problem). In these cases, a different central
engine, likely a hyper-accreting black hole, might be involved. However, from the distribution of
elements (e.g., iron or oxygen) through nebular emission lines, Mazzali et al. (2001) developed 2D
explosion models and 3D radiation transport calculations (Maeda et al. 2002; Tanaka et al. 2007), which
suggested that SNe are likely aspherical. The derived $E_{\rm SN}$ is likely overestimated, and the
real $E_{\rm SN}$ may be smaller by a factor of (2-5) (Mazzali et al. 2014). If we recalculated $E_{\rm
SN}$ by reducing the isotropic value by a factor of three, we find that the $E_{\rm SN}$ values of all
the systems are roughly in the range of $(0.2-2)\times 10^{52}$ erg (see Figure
\ref{fig:TOTALGRBSN}(b)), and the total energy of most GRB/SN system are below the energy budget of a
magnetar (see Figure \ref{fig:TOTALGRBSN}(c)). Overall, the data are consistent with-even though not a
proof of-the hypothesis that all GRB/SN systems have a magnetar central engine. This conclusion is
consistent with that of Mazzali et al. (2014).

\section{Conclusions and Discussion}

To understand the origin of GRB/SN systems, we systematically study a sample of 20 GRB/SN association
systems with robust spectroscopic evidence of the associated SNe. For comparison, we also include other
typical long GRBs without observed SN association (Amati et al. 2002; Zhang et al. 2009) and other Type
Ib/c SNe without associated GRBs (Lyman et al. 2016). By deriving/collecting basic physical parameters
of GRBs and SNe and analyzing their correlations, we are able to reach several interesting conclusions.
\begin{itemize}
 \item The peak brightness, $^{56}\rm Ni$ mass and explosion energy of the SNe in our sample are
     systematically higher than other Type Ib/c SNe without associated GRBs with a K-S test value
     $p_{KS}=0.1$. This hints that an additional energy source other than $^{56}\rm Ni$ decay might
     be playing the role to power the SNe.
 \item The beaming-corrected GRB $\gamma$-ray energy $E_{\rm \gamma}$ and kinetic energy $E_{\rm
     K}$ in our sample are both less than the maximum available energy of a millisecond magnetar.
     The SN energy $E_{\rm SN}$ of most systems is also smaller than this energy budget. When
     aspherical explosions are assumed, most SNe in our sample are below the energy budget limit of
     a magnetar. The total GRB+SN energy of most systems in our sample are below or close to the
     maximum rotation energy of a magnetar when assuming aspherical SN explosions, with the SN
     energy distributed in the range $(0.2-2)\times 10^{52}$ erg. All these are consistent (but not
     a proof) of the hypothesis that most, if not all, GRB/SN systems are powered by millisecond
     magnetars. Indeed, a few hyper-energetic GRB/SN events are identified, which may require a
     hyper-accreting black hole as the central engine.
 \item The energy partition between GRB and SN in these systems is such that most of the energy is
     carried by the SN. The GRB energy is typically less than 30\%, with a center value of about
     10\% of the total budget.
 \item Several interesting statements may be made regarding some correlations. First, even though
     most systems in our sample satisfy the so-called Amati relation ($E_{\rm \gamma, iso}-E_{\rm
     p,i}$, Amati et al. 2002), there are apparent outliers in the GRB/SN systems to this relation.
     A tentative relation between $E_{\rm p}$ and $M_{\rm peak}$ (Li 2006) is confirmed in our
     study.
\end{itemize}

\begin{figure}
\includegraphics[angle=0,scale=0.6]{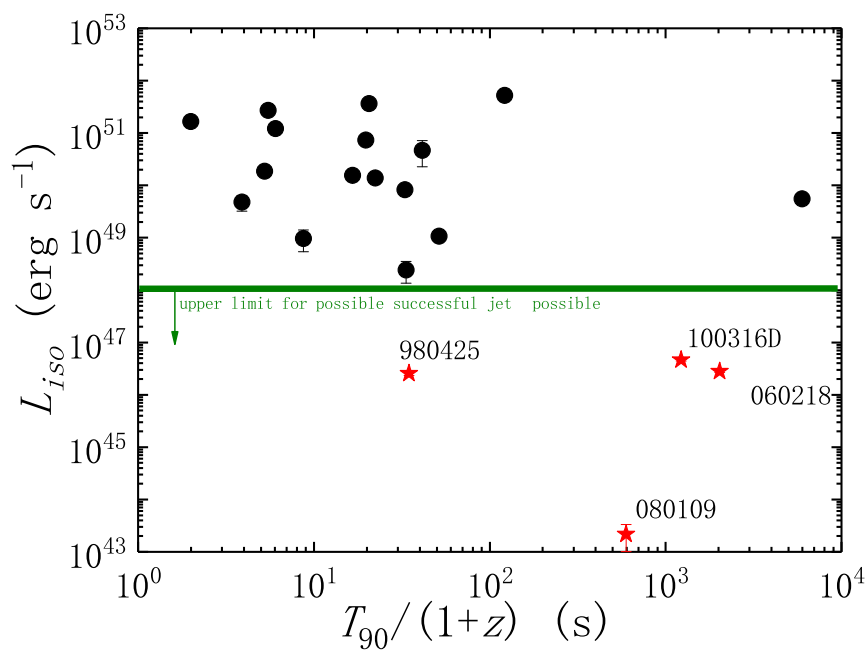}
\center\caption{Isotropic luminosity of GRBs as a function of $T_{90}/(1 + z)$ in our sample. Black dots
denote the engine-driven GRBs, while red stars denote the
possible shock-breakout GRBs suggested in the literature. The horizontal solid line ($10^{48}\rm~ erg~s^{-1}$) is
a rough threshold above which successful jet breakout is possible
(Zhang et al. 2012).}
\label{fig:LUMT90}
\end{figure}

Some famous GRB/SN associations (e.g,. GRB 980425/SN 1998bw and GRB 060218/SN 2006aj) belong to the
so-called low-luminosity GRBs (e.g., Campana et al. 2006; Liang et al. 2007; Soderberg et al. 2008;
Virgili et al. 2009). Some authors suggested that these systems may be related to shock breakouts
(e.g., Li 2007; Wang \& M\'{e}sz\'{a}ros 2007; Chevalier \& Fransson 2008; Bromberg et al. 2011; Nakar
\& Sari 2012). These bursts typically have smooth, long-duration burst light curves and low
luminosities. Successful jets typically have variable light curves and high luminosities. The
separation line between the two types of GRBs is $\sim 10^{48}~\rm erg~s^{-1}$ (Zhang et al. 2012). In
Figure \ref{fig:LUMT90}, we plot the GRB luminosity against duration in our sample. It is interesting
to see that only four systems in our sample are below the low-luminosity GRB category defined by Zhang
et al. (2012). Most other GRBs in our sample are actually high-luminosity GRBs. The similarity of the
SN properties between low-luminosity and high-luminosity GRBs suggests that they likely share a similar
type of the progenitor star, with the difference in the GRB emission properties defined by some
parameters related to jet launching (e.g., jet power, engine duration).

Besides serving as central engines in GRB/SN systems, young magnetars have been invoked to power other
systems as well, including super-luminous supernovae (e.g., Kasen \& Bildsten 2010; Woosley 2010; Wang
et al. 2015, 2016; Dong et al. 2016), NS-NS mergers (Dai et al. 2006; Fan \& Xu 2006; Metzger et al.
2008; Rowlinson et al. 2010, 2013; Gao et al. 2013a; Yu et al. 2013; Zhang 2013; Metzger \& Piro 2014;
L\"{u} et al. 2015; Gao et al. 2016), and even fast radio bursts (Zhang 2014; Murase et al. 2016;
Metzger et al. 2017). The different properties of these transient events may be related to different
parameters of the underlying magnetars (Metzger et al. 2015; Yu et al. 2017).

\acknowledgments

We thank J. D. Lyman for sharing the Type Ib/c SNe data with us, Wei-Hua Lei for helpful discussion,
and the anonymous referee for helpful comments. D.A.K acknowledges financial support from the Spanish
research project AYA 2014-58381-P, and from Juan de la Cierva Incorporaci\'on fellowship
IJCI-2015-26153. This work is supported by the National Basic Research Program (973 Programme) of China
2014CB845800, the National Natural Science Foundation of China (grant Nos.11603006, 11851304, 11533003,
and U1731239), Guangxi Science Foundation (grant Nos. 2017GXNSFFA198008, 2016GXNSFCB380005,
2015GXNSFDA39002, and AD17129006), the One-Hundred-Talents Program of Guangxi colleges, the high-level
innovation team and outstanding scholar program in Guangxi colleges, Scientific Research Foundation of
Guangxi University (grant No. XGZ150299), and special funding for Guangxi distinguished professors
(Bagui Yingcai \& Bagui Xuezhe).



\begin{center}
\begin{deluxetable}{lllllllllllll}
\tablewidth{0pt} \tabletypesize{\footnotesize} \tabletypesize{\tiny} \tablecaption{The properties of
prompt and afterglow emission of GRBs in our sample.}\tablenum{1}

\tablehead{\colhead{GRB/SN}& \colhead{Detectors\tablenotemark{a}}& \colhead{Redshift}&
\colhead{$T_{90}$}&  \colhead{$E_{p}$\tablenotemark{b}}&  \colhead{$~\rm \theta_{j}$
\tablenotemark{c}}& \colhead{References}\\\colhead{(Name)}&\colhead{}& \colhead{($z$)}&\colhead{(s)}&
\colhead{(keV)}& \colhead{(Degree)}& \colhead{}}

\startdata \hline 980425/1998bw	&	{\em BeppoSAX}    &	0.0085	&	35	        	&	55$\pm$21          	
&	11$\pm$3    	&(1)-(3)	\\
011121/2001ke	&	{\em BeppoSAX/KW}    &	0.362	&	28	   &	$819^{+108}_{-96}$	&	
4.49$\pm$0.16	        &(4)	\\
021211/2002lt	&	{\em HETE-II}	    &	1.006	&	4	   &	47$\pm$5	&		
4.82$\pm$0.68	        &(5)	\\
030329/2003dh	&	{\em HETE-II}	    &	0.1685	&	23	   &	79$\pm$3	&		
3.8$\pm$0.05	&(1),(6)	\\
031203/2003lw	&	{\em INTEGRAL}    &	0.1055	&	37	   &	159$\pm$51	         &		
9$\pm$2	        &(1),(7)	\\
050525A/2005nc	&	{\em Swift/KW}	    &	0.606	&	8.8	   &	84$\pm$2	&	2.12$\pm$0.46	&
(8),(9)	\\
060218/2006aj	&	{\em Swift}	    &	0.0334	&	2100   &	4.9$\pm$1.2	&	
12.6$\pm$3.95	&(1),(10)-(12)	\\
080109/2008d	&	{\em Swift}	    &	0.007	&	600	   &$0.12^{+0.23}_{-0.089}$ &	
$8.09\uparrow$	        &(13)-(15)	\\
081007/2008hw	&	{\em Swift/Fermi}	    &	0.5295	&	8	   &	61$\pm$15	&	$11.09\uparrow$	
&(16),(17)	\\
091127/2009nz	&	{\em Swift/KW/Fermi}   &	0.49	&	9	   &	36$\pm$2	&	5.5$\pm$1.5	
&(17)-(19)	\\
100316D/2010bh	&	{\em Swift}	    &	0.0591	&	$>1300$  &$18^{+3}_{-2}$	&	$5.6\uparrow$	
&(1),(20),(21)	\\
101219B/2010ma	&	{\em Swift/Fermi}	    &	0.55	&	51	   &	70$\pm$8	&	
$9.07\uparrow$	    &(17),(22)	\\
111209A/2011kl	&	{\em Swift/KW}	    &	0.677	&	$\sim10000$	   &	520$\pm$89	&	
9.17$\pm$1.5	    &(23)-(25)	\\
120422A/2012bz	&	{\em Swift}	    &	0.283	&	5	   &$33^{+39}_{-33}$	&	23$\pm$7	
&(26)-(28)	\\
130215A/2013ez	&	{\em Swift/Fermi}	    &	0.597	&	66	   &	155$\pm$63	&	$10.16\uparrow$	
&(29)	\\
130427A/2013cq	&	{\em Fermi/Swift}	&	0.3399	&	163	   &	830$\pm$5	&	$7.5\uparrow$	
&(30)-(33)	\\
130702A/2013dx	&	{\em Fermi/KW}	    &	0.145	&	59	   &	10$\pm$1	&14$\pm$4	
&(34)-(36)	\\
130831A/2013fu	&	{\em Swift/KW}	    &	0.479	&	33	   &	67$\pm$4	&	
$3.17\uparrow$	    &(29)	\\
140606B/iPTF14bfu	&	{\em Fermi/KW}	    &	0.384	&	 23	   &	 579$\pm$135	&	
$11.5\uparrow$	
&(37)	\\
161219B/2016jca	&	{\em Swift/KW}	&	0.1475	&	10	&	93$\pm$29		&	$\sim 40$	&(38),(39)	
\\ 		 																	
\enddata

\tablenotetext{a}{Detected by different instruments: KW is Konus-Wind.} \tablenotetext{b}{The peak
energy in the $E^2\rm{N}(E)$ spectrum of the prompt emission.} \tablenotetext{c}{The jet opening angle
of GRBs measured from the afterglow. Upward-pointing arrows denote lower limits for the jet opening
angles.}

\tablerefs{(1)Hjorth \& Bloom 2012;(2)Iwamoto 1999;(3)Kouveliotou et al. 2004;(4)Tsvetkova et al.
2017;(5)Della Valle et al. 2003;(6)Deng et al. 2005(7)Gal-Yam et al. 2004;(8)Della Valle et al.
2006;(9)Kovacevic et al. 2014;(10)Ferrero et al. 2006;(11)Campana et al. 2006;(12)Mirabal et al.
2006;(13)Xu et al.2008;(14)Li 2008;(15)Soderberg et al. 2008;(16)Jin et al. 2013;(17)Olivares E et al.
2015;(18)Berger et al. 2011;(19)Vergani et al. 2011;(20)Bufano et al. 2012;(21)Fan et al.
2011;(22)Larsson et al. 2015;(23)Greiner et al. 2015;(24)Kann et al. 2016;(25)Kann et al.
2017;(26)Zhang et al. 2012;(27)Melandri et al. 2012;(28)Schulze et al. 2014;(29)Cano et al. 2014;(30)Xu
et al.2013;(31)Vestrand et al. 2014;(32)Ackermann et al. 2014;(33)Maselli et al. 2014;(34)Singer et al.
2013;(35)D'Elia et al. 2015;(36)Volnova et al. 2017;(37)Cano et al. 2015;(38)Ashall et al.
2017;(39)Cano et al. 2017a}

\end{deluxetable}
\end{center}


\begin{center}
\begin{deluxetable}{lllllllllllll}
\tablewidth{0pt} \tabletypesize{\footnotesize} \tabletypesize{\tiny} \tablecaption{The observational
properties and derived parameters of SNe associated with GRBs.}\tablenum{2}

\tablehead{\colhead{GRB/SN}& \colhead{SN}& \colhead{SN\tablenotemark{a}}& \colhead{$M_{\rm
peak}$\tablenotemark{b}}& \colhead{$t_{\rm peak}$\tablenotemark{b}}&
 \colhead{$M_{\rm Ni}$\tablenotemark{c}}& \colhead{$ M_{\rm ejec}$\tablenotemark{d}}&
\colhead{References}\\\colhead{(Name)}&\colhead{(Type)}& \colhead{(Evidence)}&\colhead{(Mag)}&
\colhead{(Day)}&\colhead{($M_{\odot}$)}& \colhead{($M_{\odot}$)}& \colhead{}}

\startdata \hline 980425/1998bw	&	Ic	&	Spec.	&	-18.86$\pm$0.2	&	$\sim17$		
&	0.54$\pm$0.02	&	6.8$\pm$0.57	&(1)-(4)	\\
011121/2001ke	&	Ic	&	Bump/spec.	&	-18.55$\pm$0.55	&	13$\pm$1		
&		0.35$\pm$0.01		&		4.44$\pm$0.82		&(5)	\\
021211/2002lt	&	Ic	&	Spec.	&	-18.8$\pm$0.4	&	$\sim14$		&		0.4$\pm$0.14		
&		7.16$\pm$5.99		&(6)	\\
030329/2003dh	&	Ic	&	Spec.	&	-18.79$\pm$0.23	&	11.5$\pm$1.5		&	
0.54$\pm$0.13	&	5.06$\pm$1.65	&(1),(7),(8)	\\
031203/2003lw	&	Ic	&	Spec.	&	-18.92$\pm$0.2	&	21.5$\pm$3.5		&	
0.57$\pm$0.04	&	8.22$\pm$0.76	&(1),(9)	\\
050525A/2005nc	&	Ic	&	Spec.	&	-18.8$\pm$0.6	&	$\sim12$		&		0.42$\pm$0.02		
&		4.75$\pm$1.08		&(10),(11)	\\
060218/2006aj	&	Ic&	Spec.	&	-18.16$\pm$0.1	&	10$\pm$0.5		&	0.28$\pm$0.08	
&	2.58$\pm$0.55	&(1),(12)-(14)	\\
080109/2008d	    &	Ib	&	Spec.	&	-16.9$\pm$0.2	&	19$\pm$0.8		&	0.09$\pm$0.01	&	
5.3$\pm$1	&(15)-(18)	\\
081007/2008hw	&	Ic	&	Bump/spec.	&	-18.5$\pm$0.5	&	12$\pm$3		&	0.39$\pm$0.06	&	
2.3$\pm$1	&(19),(20)	\\
091127/2009nz	&	Ic	&	Bump/spec.	&	-18.65$\pm$0.2	&	15$\pm$2		&	0.33$\pm$0.01	
&	4.69$\pm$0.13	&(20)-(22)	\\
100316D/2010bh	&	Ic	&	Spec.	&	-18.45$\pm$0.18	&	8.48$\pm$1.06		&	0.12$\pm$0.02	&	
2.47$\pm$0.23	
&(1),(23)	\\
101219B/2010ma	&	Ic	&	Spec.	&	-18.5$\pm$0.25	&	10$\pm$2		&	0.43$\pm$0.03	
&	1.3$\pm$0.4	&(20),(24)	\\
111209A/2011kl	&	SLSN	&	Spec.	&	-19.8$\pm$0.2	&14$\pm$0.5		&	1$\pm$0.1	&	
3$\pm$1	&(25)-(27)	\\
120422A/2012bz	&	Ic&	Spec.	&	-18.56$\pm$0.15	&	16.69$\pm$1.28		&	
0.57$\pm$0.07	&	6.1$\pm$0.49	&(28)-(30)	\\
130215A/2013ez	&	Ic	&	Spec.	&	-18.85$\pm$0.15	&	
6.41$\pm$0.34	&	0.375$\pm$0.025	&		---		&(31)	\\
130427A/2013cq	&	Ic	&	Spec.	&	-18.91$\pm$0.2	&	$\sim15.2$		&	0.38$\pm$0.02	
&	6.27$\pm$0.69	&(32)-(35)	\\
130702A/2013dx	&	Ic	&	Spec.	&	-18.4$\pm$0.4	&	17.2$\pm$0.34	&	
	0.38$\pm$0.01	&		3$\pm$0.1		&(36)-(38)	\\
130831A/2013fu	&	Ic&	Spec.	&	-18.89$\pm$0.05	&	18.53$\pm$0.07	&	
	0.48$\pm$0.07	&	6.71$\pm$0.2	&(31)	\\	

140606B/iPTF14bfu	&	Ic&	Spec.	&	-19.61$\pm$0.27	&	16.32$\pm$1.63	&	0.4$\pm$0.2	&	5$\pm$2	
&(39)	\\	
																											
161219B/2016jca	&	Ic&	Spec.	&	-19.04$\pm$0.05	&	10.7$\pm$0.3		&	
0.4$\pm$0.1	&	5.8$\pm$0.3	&(40),(41)	\\
\enddata
\tablenotetext{a}{The evidence of a SN associated with a GRB. ``spec.'' is strong spectroscopic
evidence, and ``bump'' is a clear light curve bump with some spectroscopic evidence.}
\tablenotetext{b}{The peak magnitute and peak time in the SN light curve.}  \tablenotetext{c}{The mass
of Nickel measured from the SN.} \tablenotetext{d}{The mass of ejecta in the blastwave.}

\tablerefs{(1)Hjorth \& Bloom 2012;(2)Weiler et al.2001;(3)Nakamura etal. 2001;(4)Clocchiatti et al.
2011;(5)Tsvetkova et al. 2017;(6)Della Valle et al. 2003;(7)Deng et al. 2005;(8)Mazzali et al.
2003;(9)Mazzali et al. 2006;(10)Della Valle et al. 2006;(11)Kovacevic et al. 2014;(12)Campana et al.
2006;(13)Mirabal et al. 2006;(14)Li 2007;(15)Xu et al. 2008;(16)Mazzali et al.2008;(17)Li
2008;(18)Soderberg et al. 2008;(19)Jin et al. 2013;(20)Olivares E et al. 2015;(21)Berger et al.
2011;(22)Cobb et al. 2010;(23)Bufano et al. 2012;(24)Sparre et al. 2011;(25)Greiner et al.
2015;(26)Kann et al. 2016;(27)Kann et al. 2017;(28)Zhang et al. 2012;(29)Melandri et al.
2012;(30)Schulze et al. 2014;(31)Cano et al. 2014;(32)Xu et al. 2013;(33)Vestrand et al.
2014;(34)Ackermann et al. 2014;(35)Melandri et al. 2014;(36)D'Elia et al. 2015;(37)Toy et al.
2016;(38)Volnova et al. 2017;(39)Cano et al. 2015;(40)Ashall et al. 2017;(41)Cano et al. 2017a}

\end{deluxetable}
\end{center}

\begin{center}
\begin{deluxetable}{lllllllllllll}
\rotate
\tablewidth{0pt} \tabletypesize{\footnotesize} \tabletypesize{\tiny} \tablecaption{The derived
energies of of GRBs and SNe in our sample.}\tablenum{3}

\tablehead{\colhead{GRB/SN}& \colhead{$~\rm E_{\gamma, iso}$\tablenotemark{a}}& \colhead{$~\rm
E_{\gamma}$\tablenotemark{a}}& \colhead{$~\rm E_{K, iso}$\tablenotemark{b}}& \colhead{$~\rm
E_{K}$\tablenotemark{b}}& \colhead{$~\rm E_{GRB, iso}$\tablenotemark{c}}& \colhead{$~\rm
E_{GRB}$\tablenotemark{c}}& \colhead{$~\rm E_{SN}$\tablenotemark{d}}& \colhead{$\eta
\%$\tablenotemark{e}}\\\colhead{(Name)}&\colhead{($\rm erg$)}& \colhead{($\rm erg$)}&\colhead{($\rm
erg$)}& \colhead{($\rm erg$)}&\colhead{($\rm erg$)}& \colhead{($\rm erg$)}&\colhead{($\rm erg$)}&
\colhead{}}

\startdata \hline

980425/1998bw	& (9$\pm$0.87)$\times10^{47}$     	&	(1.66$\pm$1.19)$\times10^{46}$    &	
$2^{+4.4}_{-0.6}\times10^{49}$	& $3.68^{+10.41}_{-1.18}\times10^{47}$	&	
$2.09^{+4.41}_{-0.61}\times10^{49}$ &	$3.85^{+10.51}_{-3.51}\times10^{47}$ &
$(1.3\pm0.1)\times10^{52}$	&	$(2.96\pm1)\times 10^{-3}$     \\

011121/2001ke	& (1.02$\pm$0.15)$\times10^{53}$     	&	(2.25$\pm$0.32)$\times10^{47}$    &	
$5.83^{+12.83}_{-1.3}\times10^{53}$	& $2.14^{+4.49}_{-0.54}\times10^{51}$	&	
$6.85^{+12.98}_{-1.45}\times10^{53}$ &	$2.51^{+5.27}_{-0.63}\times10^{51}$ &
$(1.77\pm0.88)\times10^{52}$	&	$12.36\pm3.64$     \\

021211/2002lt	& (6.6$\pm$0.6)$\times10^{51}$     	&	(2.33$\pm$0.92)$\times10^{49}$    &	
$6.62^{+14.58}_{-1.46}\times 10^{52}$	&  $2.34^{+5.86}_{-0.19}\times10^{50}$	&	
$7.28^{+14.61}_{-1.52}\times10^{52}$ &	$2.57^{+5.93}_{-1.31}\times10^{50}$ &
$(2.85\pm1.3)\times10^{52}$	&	$0.89\pm0.45$     \\

030329/2003dh	& (1.7$\pm$0.3)$\times10^{52}$     	&	(3.74$\pm$3.56)$\times10^{49}$    &	
$1.82^{+4.1}_{-0.4}\times10^{52}$	&  $4^{+14.31}_{-2.44}\times10^{49}$	&	
$3.52^{+4.4}_{-0.7}\times10^{52}$ &	$7.73^{+2}_{-1.18}\times10^{49}$ &
$(1.21\pm0.39)\times10^{52}$	&	$0.63\pm0.27$     \\

031203/2003lw	& (9$\pm$4)$\times10^{49}$     	&	(1.11$\pm$1.04)$\times10^{48}$    &	
$1.44^{+3.16}_{-0.32}\times10^{51}$	&  $1.77^{+4.77}_{-0.48}\times10^{49}$	&	
$1.53^{+3.2}_{-0.36}\times10^{51}$ &	$1.89^{+4.88}_{-1.37}\times10^{49}$ &
$(1.59\pm0.15)\times10^{52}$	&	$0.12\pm0.35$     \\

050525A/2005nc	& $(2.39\pm0.15)\times10^{52}$     	&	(1.57$\pm$0.76)$\times10^{49}$    &	
$4.48^{+9.85}_{-0.98}\times10^{52}$	&  $3.06^{+8.21}_{-0.81}\times10^{49}$	&	
$6.78^{+9.85}_{-0.98}\times10^{52}$ &	$4.64^{+8.97}_{-2.9}\times10^{49}$ &
$(1.89\pm0.75)\times10^{52}$	&	$0.25\pm0.49$     \\

060218/2006aj	& (5.9$\pm$0.3)$\times10^{49}$     	&	(1.43$\pm$1.11)$\times10^{48}$    &	
$2.67^{+5.88}_{-0.59}\times10^{49}$	&  $6.45^{+18.9}_{-3.25}\times10^{47}$	&	
$8.57^{+6.18}_{-0.89}\times10^{49}$ &	$2.07^{+2.99}_{-1.72}\times10^{48}$ &
$(6.1\pm0.14)\times10^{51}$	&	$0.03\pm0.05$    \\

080109/2008d	& ($1.3^{+1.5}_{-0.7}$)$\times10^{46}$     	&	1.3$\times10^{44}\uparrow$    &	
$4.46^{+9.84}_{-0.99}\times10^{48}$ 	& 4.44$\times10^{46}\uparrow$	&	
$4.47^{+9.84}_{-0.98}\times10^{48}$ &	4.45$\times10^{46}\uparrow$ &
$(6\pm3)\times10^{51}$	&	$7.42\times 10^{-4}\uparrow$    \\

081007/2008hw	& $\sim1.5\times10^{51}$     	&	2.81$\times10^{49}\uparrow$    &	
$7.52^{+17.08}_{-1.66}\times10^{52}$ 	& 1.41$\times10^{51}\uparrow$	&	
$7.67^{+17.71}_{-1.65}\times10^{52}$ &	1.43$\times10^{51}\uparrow$ &
$(9\pm5)\times10^{51}$	&	$13.7\uparrow$     \\

091127/2009nz	& $\sim1.1\times10^{52}$     	&	(5.06$\pm$3.14)$\times10^{49}$    &	
$3.33^{+7.32}_{-0.74}\times10^{52}$	&  $1.53^{+4.34}_{-0.61}\times10^{50}$	&	
$4.43^{+7.37}_{-0.73}\times10^{52}$ &	$2.04^{+4.66}_{-1.6}\times10^{50}$ &
$(8.1\pm0.2)\times10^{51}$	&	$2.45\pm5.81$    \\

100316D/2010bh	& (6$\pm$0.3)$\times10^{49}$     	&	2.86$\times10^{47}\uparrow$    &	
$5.88^{+12.94}_{-1.3}\times10^{53}$ 	& 2.8$\times10^{51}\uparrow$	&	
$5.88^{+12.94}_{-1.3}\times10^{53}$ &	2.8$\times10^{51}\uparrow$ &
$(9.2\pm0.8)\times10^{51}$	&	$23.33\uparrow$     \\

101219B/2010ma	& (4.2$\pm$0.3)$\times10^{51}$     	&	5.26$\times10^{49}\uparrow$    &	
$1.76^{+3.88}_{-0.39}\times10^{53}$ 	& 2.2$\times10^{51}\uparrow$	&	
$1.8^{+13.8}_{-0.39}\times10^{53}$ &	2.25$\times10^{51}\uparrow$ &
$(1\pm0.6)\times10^{52}$	&	$18.37\uparrow$     \\

111209A/2011kl	& (5.54$\pm$0.7)$\times10^{53}$     	&	(7.09$\pm$3.4)$\times10^{51}$    &	
$8^{+17.6}_{-1.76}\times10^{53}$	&  $1.02^{+2.61}_{-0.14}\times10^{52}$	&	
$1.35^{+1.83}_{-0.25}\times10^{54}$ &	$1.73^{+2.95}_{-0.93}\times10^{52}$ &
$(5.5\pm3.5)\times10^{51}$	&	$75.87\pm16.6$     \\

120422A/2012bz	& (2.4$\pm$0.8)$\times10^{50}$     	&	(1.93$\pm$1.2)$\times10^{49}$    &	
$2.94^{+6.49}_{-0.65}\times10^{52}$	&  $2.37^{+6.88}_{-1.14}\times10^{51}$	&	
$2.97^{+6.49}_{-0.66}\times10^{52}$ &	$2.39^{+6.9}_{-2.21}\times10^{51}$ &
$(1.53\pm0.13)\times10^{52}$	&	$13.5\pm14.3$     \\

130215A/2013ez	& ($3.1^{+0.9}_{-1.6}$)$\times10^{52}$     	&	4.87$\times10^{50}\uparrow$    &	
$1.23^{+8.1}_{-0.26}\times10^{53}$ 	& 1.93$\times10^{51}\uparrow$	&	
$1.54^{+2.7}_{-0.27}\times10^{53}$ &	2.42$\times10^{51}\uparrow$ &
---	&	--     \\

130427A/2013cq	& (8.5$\pm$0.04)$\times10^{53}$     	&	7.17$\times10^{51}\uparrow$    &	
$3.0^{+11.82}_{-1.19}\times10^{53}$	&  2.54$\times10^{51}\uparrow$	&	
$1.5^{+1.19}_{-0.12}\times10^{54}$ &	9.71$\times10^{51}\uparrow$ &
$(6.39\pm0.7)\times10^{52}$	&	$13.2\uparrow$      \\

130702A/2013dx	& (6.36$\pm$1.34)$\times10^{50}$     	&	(1.9$\pm$1.64)$\times10^{49}$    &	
$1.92^{+5.23}_{-0.43}\times10^{52}$	&  $5.73^{+16.4}_{-2.49}\times10^{50}$	&	
$1.98^{+4.24}_{-0.43}\times10^{52}$ &	$5.9^{+16.5}_{-5.15}\times10^{50}$ &
$(8.2\pm0.4)\times10^{51}$	&	$6.71\pm8.5$    \\

130831A/2013fu	& (4.6$\pm$0.2)$\times10^{51}$     	&	7.06$\times10^{48}\uparrow$    &	
$5.27^{+11.59}_{1.16}\times10^{53}$ 	& 8.08$\times10^{50}\uparrow$	&	
$5.32^{+11.64}_{-1.16}\times10^{53}$ &	8.16$\times10^{50}\uparrow$ &
$1.87_{-0.62}^{+0.9}\times10^{52}$	&	$4.18\uparrow$     \\

140606B/iPTF14bfu	& (3.47$\pm$0.02)$\times10^{51}$     	&	6.94$\times10^{49}\uparrow$    &	
$4.2\pm1.4\times10^{52}$ 	& 8.4$\times10^{50}\uparrow$	&	$4.5\pm1.4\times10^{52}$ &	
9.1$\times10^{50}\uparrow$ &
$2\pm1\times10^{52}$	&	$4.31\uparrow$     \\

161219B/2016jca	& (9.7$\pm$4.3)$\times10^{49}$    	&	(2.36$\pm$1.05)$\times10^{49}$    &	
$1.6^{+3.52}_{-0.36}\times10^{50}$	&  $3.9^{+8.57}_{-0.85}\times10^{49}$	&	
$2.57^{+3.95}_{-0.78}\times10^{50}$ &	$6.29^{+9.62}_{-2.69}\times10^{49}$ &
$(5.1\pm0.8)\times10^{52}$	&	$0.12\pm0.24$     \\
\enddata

\tablenotetext{a}{The isotropic and jet-corrected prompt $\gamma$-ray emission energy of GRBs is
calculated by using fluence and redshift extrapolated into 1-10,000 keV (rest frame) with a spectral
model and a k-correction. The value of GRBs 011121 and 050525A are taken from Kan et al. 2010}
\tablenotetext{b}{The isotropic and jet-corrected kinetic energy of GRBs measured from the afterglow
flux during the normal decay phase.} \tablenotetext{c}{The isotropic and jet-corrected total energy of
GRBs.} \tablenotetext{d}{The isotropic SN energy.} \tablenotetext{e}{The efficiency of GRB/SN events.}

\end{deluxetable}
\end{center}

\clearpage


\end{document}